\documentclass[draft]{agujournal2018}
\usepackage{url} 
\usepackage{soul}
\usepackage{subscript}
\usepackage{booktabs}

\usepackage{textcomp} 
\usepackage{siunitx}  
\usepackage{subfig}
\usepackage{comment}
\definecolor{DarkGreen}{rgb}{0.0,0.70,0.0}

\usepackage{natbib} 
\bibpunct{(}{)}{;}{a}{,}{,~}

\usepackage{enumitem} 
\usepackage{float} 
\usepackage{setspace}
\usepackage{hyperref}
\hypersetup{colorlinks=true,citecolor=black,filecolor=black,linkcolor=black,urlcolor=black,breaklinks=true}

\newcommand\blfootnote[1]{%
	\begingroup
	\renewcommand\thefootnote{}\footnote{#1}%
	\addtocounter{footnote}{-1}%
	\endgroup
}

\journalname{Journal of Geophysical Research: Biogeosciences}

\begin{document}

%
%

\title{Land Cover Changes Cause Increased Losses during Photosynthetic Extremes}

\blfootnote{This manuscript has been authored by UT-Battelle, LLC, under contract DE-AC05-00OR22725 with the US Department of Energy (DOE). The US government retains and the publisher, by accepting the article for publication, acknowledges that the US government retains a nonexclusive, paid-up, irrevocable, worldwide license to publish or reproduce the published form of this manuscript, or allow others to do so, for US government purposes. DOE will provide public access to these results of federally sponsored research in accordance with the DOE Public Access Plan (http://energy.gov/downloads/doe-public-access-plan).}

%
%




\authors{Bharat~Sharma\affil{1,2}, Jitendra~Kumar\affil{3}, Nathan~Collier\affil{2}, and Auroop~R.~Ganguly\affil{1}, Forrest~M.~Hoffman\affil{2,4}}

\affiliation{1}{Sustainability and Data Sciences Laboratory, Department of Civil and Environmental Engineering, Northeastern University, Boston, Massachusetts, USA}
\affiliation{2}{Computational Sciences \& Engineering Division and the Climate Change Science Institute, Oak Ridge National Laboratory, Oak Ridge, Tennessee, USA}
\affiliation{3}{Environmental Sciences Division, Oak Ridge National Laboratory, Oak Ridge, Tennessee, USA}
\affiliation{4}{Department of Civil and Environmental Engineering, University of Tennessee, Knoxville, Tennessee, USA}





\correspondingauthor{Bharat Sharma}{bharat.sharma.neu@gmail.com}




\begin{keypoints}
\item Human activities, through land-use change, lead to increased intensity, duration, and frequency of negative extremes in GPP.
\item Precipitation anomaly is the dominant trigger for GPP extremes, while soil moisture anomaly leads to extended extreme events.
\item The regions with overall reduction in GPP often show weakening of negative extremes in GPP.
%
%
\end{keypoints}

\begin{abstract}

Human-induced carbon dioxide (CO\textsubscript{2}) emissions, primarily from fossil fuel combustion and changes in land use and land cover (LULCC), are a key contributor to climate change. As the climate warms, extreme events such as heatwaves, droughts, and wildfires have become more frequent and are projected to intensify throughout the 21st century. These escalating extremes are likely to further disrupt vegetation productivity, known as gross primary production (GPP), and reduce the ecosystem's capacity to absorb carbon. In this study, we employ a global Earth system model to assess how (a) CO\textsubscript{2} emissions alone and (b) CO\textsubscript{2} combined with LULCC influence the severity, frequency, and duration of GPP extremes. Our results show that negative GPP extremes periods of unexpectedly low carbon uptake are increasing more rapidly than positive extremes, especially under LULCC scenarios. The primary climate factor driving these extremes is soil moisture variability, which is influenced by fluctuations in both precipitation and temperature. The delayed responses of GPP to different climate drivers depend on the specific driver and geographical region. Overall, the highest incidence of GPP extremes arises from the combined influence of water stress, temperature anomalies, and fire-related disturbances.

\end{abstract}

%
%
%


\section*{Plain Language Summary}

\noindent Rising carbon dioxide (CO\textsubscript{2}) emissions due to human activities, such as fossil fuel burning and land use and land cover change (LULCC), are the major driver of climate change. Heatwaves, droughts, and fires have increased and are expected to accelerate with climate change in the twenty-first century and beyond. This increase in extreme climate conditions has the potential to further alter vegetation productivity (called gross primary production or GPP) and carbon uptake capacity. Here, we use a global Earth system model to investigate the impacts of (1) CO\textsubscript{2} forcing and (2) CO\textsubscript{2} and LULCC forcing on the intensity, frequency, and duration of extreme events in GPP. We found that the negative extremes in GPP, which are associated with higher losses in carbon uptake than expected, increase at a higher rate than positive GPP extremes; and this rate rises with LULCC forcing. The most dominant climate driver causing the GPP extremes is soil moisture anomalies, which are triggered by extremes in precipitation and temperature. The lagged responses of climate drivers on GPP extremes vary with the drivers and spatial location. The largest number of GPP extremes were driven by the compound effect of water-, temperature-, and fire-related drivers. 

%
%

%


%
%
%
%


\section{Introduction}
\label{sec:intro}

Human activities are altering the Earth's atmosphere, ocean, and land surfaces at a scale and magnitude not seen throughout the past multiple thousands of years.
The rising concentration of greenhouse gases (GHGs) such as water vapor, ozone, carbon dioxide, methane, and nitrous oxide are primary drivers of global warming and climate change.
As a result of continued growth in the global population, the demand for fossil fuels, crops, and wood is increasing.
Enhanced emissions through use of fossil fuels \citep{LeQuere_GCB_2018} and land use and land cover change (LULCC) increases the atmospheric concentration of carbon dioxide (CO\textsubscript{2}).
Rising CO\textsubscript{2} emissions have led to increased climate variability and frequency of climate extremes, which have a large impact on terrestrial gross primary productivity (GPP) and GPP extremes \citep{Marcolla_2020_NBP_Control, Xian_2020_LULCC_USA, Reichstein_2013_climate_ext, Frank_ClimateExtremes_CC_2015, Ichii_IAVgpp_2005, Piao_2019_IAV_carbon_cycle}.
While climate extremes are relatively easy to measure, their impact on terrestrial vegetation is less detectable \citep{Zscheischler_GRL_2014}.
Since terrestrial ecosystems absorb a quarter of anthropogenic CO\textsubscript{2} emissions \citep{LeQuere_GCB_2018}, large changes in vegetation productivity could alter the global carbon budget.  
Hence, it is crucial to investigate extreme anomalies in terrestrial carbon cycle and estimate their impacts on the terrestrial carbon sink under rising CO\textsubscript{2} emissions and LULCC scenarios.

Prior studies \citep{Zhu_2016_Greening, Reichstein_2013_climate_ext, Bonan_Book_2015} have found that the combined effect of CO\textsubscript{2} fertilization, increasing temperature, nitrogen deposition, and lengthening of growing seasons lead to increased vegetation productivity and strengthening of carbon sinks with a negative feedback to climate change.
However, coupled carbon-climate models have large uncertainties in future projections of ecosystem responses and feedback strength \citep{Hoffman_2014_CCTM, Reichstein_2013_climate_ext, Frank_ClimateExtremes_CC_2015,Ichii_IAVgpp_2005}.
\citet{Hoffman_2014_CCTM} found persistent atmospheric CO\textsubscript{2} biases in Coupled Model Intercomparison Project~5 (CMIP5) models because of uncertainties in biological and physical processes related to carbon accumulation.
While most Earth system model-based climate change studies analyze projections till year 2100, these projections may miss physical-biogeochemical feedbacks that arise later from the cumulative effects of climate warming \citep{Moore_Science_2018}.
The negative sensitivity of terrestrial carbon cycle to rising temperature will likely have increasing adverse implications on carbon cycle extremes over time \citep{Hubau_2020_asynchronous, Frank_ClimateExtremes_CC_2015}. 
Understanding the direction and strength of these feedbacks is essential for estimating long-term CO\textsubscript{2} concentrations and predicting and mitigating the impact and extent of climate change.
These limitations could alter the assessment of the rate of increase of atmospheric CO\textsubscript{2} and intensity of associated feedbacks with the terrestrial biosphere.

While the effects of increased warming due to greenhouse gases are spatially extensive, the LULCC effects are more regional \citep{Pitman_lulccTempPrcp_2012}.
The land-use history reconstruction used in this study estimated the proportion of land surface impacted by human activities to be 60\% of total vegetated area, mainly due to conversions from primary vegetation to managed vegetation by 2100 \citep{Hurtt_2011_Clim_Change_LULCC}.
The conversion of land from natural to managed ecosystems reduces the carbon sink and its capacity to uptake anthropogenic CO\textsubscript{2} and influences climate by modifying biogeophysical and biogeochemical processes \citep{Bonan_2012_LULCC}.
Changes in the plant functional type (PFT) at any grid cell modifies the distribution of above and below ground carbon \citep{Oleson_CLM4_2010}, response to light and energy \citep{Bonan_2011_ConopyProcesses}, distribution of soil organic matter and nutrients \citep{Koven_2013_CN}.
Human activities, such as the conversion of forests and grasslands to agricultural land and urbanization, alter net radiation, sensible and latent heat partitioning, biogeochemical cycles, and the hydrologic cycle.
Reduction of temperate vegetation cover by deforestation increases the albedo of the surface which decreases the net radiation that drives surface cooling and reduces evapotranspiration that may result in declines in precipitation; but tropical deforestation for pastures decreases the total atmospheric column heating and atmospheric vertical motion which leads to a warmer and drier climate \citep{Bonan_2012_LULCC, Bonan_Book_2015}. 
Since interannual variability (IAV) in GPP is strongly influenced by interannual variations in radiation, temperature, and precipitation \citep{Ichii_IAVgpp_2005}, the impact of LULCC in addition to CO\textsubscript{2} on carbon cycle extremes will likely increase over time.

Recent studies have investigated the characteristics of extreme anomalies in GPP due to climate change until the year 2100 \citep{Zscheischler_GRL_2014, Frank_ClimateExtremes_CC_2015, Flach_Climate_extreme_GPP_2020, Reichstein_2013_climate_ext} and a few concluded that losses in carbon uptake due to negative extremes in GPP are compensated by increased CO\textsubscript{2} fertilization \citep{Zscheischler_GRL_2014, Reichstein_2013_climate_ext}.
However, to our knowledge, no study has examined the extreme anomalies in carbon cycle beyond 2100 and the role of human LULCC in modifying carbon cycle extremes.

Rising CO\textsubscript{2} and LULCC impacts many components of terrestrial carbon cycle, namely total ecosystem carbon, net biome productivity, net ecosystem productivity, net primary productivity, and GPP.
The overarching goal of this study is to investigate the role of
CO\textsubscript{2} and LULCC in modifying the characteristics of
extremes in one of the components, Gross Primary Productivity (GPP), and attribute changes to individual and compound effects of climate drivers.
We hypothesize that 
1) rising CO\textsubscript{2} emissions will lead to larger increases in the intensity, frequency and duration of negative carbon cycle extremes than positive extremes; and 
2) LULCC activities in addition to CO\textsubscript{2} emissions will further increase GPP interannual variability and magnitude of carbon cycle extremes though total GPP will reduce.
We performed analysis to 
1) examine the magnitude, duration, frequency and spatial distribution of negative and positive carbon cycle extremes;
2) investigate the lagged response of antecedent climate drivers (and their compound effect) that drive carbon cycle extremes;
3) analyze the climate conditions that trigger long duration temporally continuous extremes;
and 4) inspect regional changes in climate-carbon feedbacks for the Central and South American tropics.

\section{Data}
\begin{figure}
 \centering
 \includegraphics[trim=1.0cm 2.0cm 1.5cm 2.0cm,width=\columnwidth]{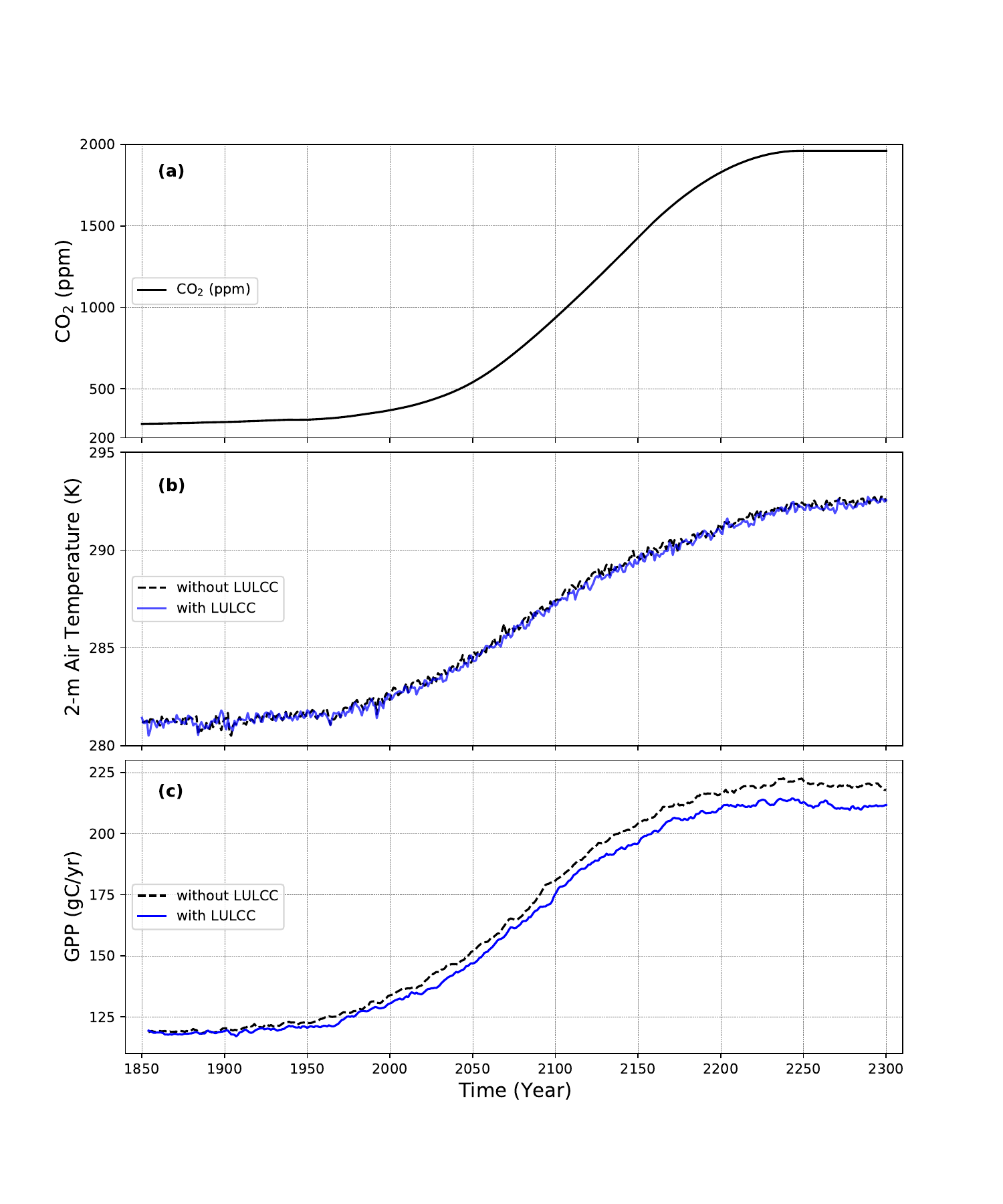}
 \caption{The (a) prescribed trajectory for atmospheric carbon dioxide (CO$_2$) forcing, (b) 5-year running mean of annual 2~m air temperature, and (c) 5-year running mean of total annual gross primary production (GPP) for the historical, RCP~8.5 and ECP~8.5 simulations.}
\label{f:co2_traj}
\end{figure}

We used simulations of the Community Earth System Model (version 1) with biogeochemistry enabled, CESM1(BGC), at approximately $1^{\circ} \times 1{^\circ}$ resolution to analyze climate-driven extreme anomalies in total photosynthetic activity.
CESM1(BGC) is a fully coupled global climate model composed of land, atmosphere, and ocean components \citep{Lindsay_2014_JClim_carbon_cycle_cesm, Lawrence_2012_lulcc_cesm1bgc}.
The atmospheric CO\textsubscript{2} forcing time series consisted of the historical (1850--2005), Representative Concentration Pathway~8.5 (RCP~8.5; 2006--2100), and Extended Concentration Pathway (ECP~8.5; 2101--2300) mole fractions, which increased from 285~ppm in 1850 to 1962~ppm in 2300 (Figure~\ref{f:co2_traj}(a)).
Analysis of the Coupled Model Intercomparison Project 5 (CMIP5) models using the International Land Model Benchmarking (ILAMB) \citep{Nate_2018_ilamb} show (Table~\ref{t:table_ilamb}) that CESM1(BGC) is one of the best performing model in terms of overall and IAV benchmark scores when compared to observational benchmarks. 

We analyzed terrestrial carbon cycle extremes (or GPP extremes) using
two simulations, namely, \emph{with} and \emph{without LULCC}.
In the simulation \emph{without LULCC}, the land cover was fixed at pre-industrial (year 1850) values.
In the \emph{with LULCC} simulation, transient land cover was prescribed over the historical and RCP~8.5 period (1850--2100) (\citet{Hurtt_2011_Clim_Change_LULCC}) and consists of the prescribed spatial distribution of PFTs (\citet{Lawrence_2012_lulcc_cesm1bgc}).
Land-use conversion is assumed to stop at the year 2100, and the distribution of PFTs remains constant at year 2100 level for the period 2100 -- 2300, while wood harvest is maintained at a constant rate over that period \citep{Mahowald_2017_GBC_LULCC}. 
The reactive nitrogen deposition followed the spatially variable time series from 1850 to 2100 \citep{Lamarque_2010_Nitrogen} and was subsequently held constant from 2100 to 2300.
Biogeochemical processes on land and in the ocean respond to the historical and prescribed RCP~8.5 and ECP~8.5 atmospheric CO\textsubscript{2} concentration forcing (Figure~\ref{f:co2_traj}(a)).
Increasing CO\textsubscript{2} fertilization, water-use efficiency, and lengthening of growing seasons enhance total photosynthesis and gross primary production (GPP).
These processes will be further altered by the changes in PFT distribution under \emph{with LULCC}.
Figure~\ref{f:co2_traj}(c) shows the 5~year moving average of global annually integrated GPP for simulations \emph{with} and \emph{without LULCC}, both forced with the same atmospheric CO\textsubscript{2} concentration trajectory (Figure~\ref{f:co2_traj}(a)).
The magnitude of the average GPP \emph{with LULCC} is less than the average GPP \emph{without LULCC} potentially due to the conversion of primary vegetation to managed vegetation.

\section{Methods}
\label{sec:methods_begin}

A significant deviation from the mean value is called an extreme.
Extreme events in GPP signify large variations in photosynthetic activity, with positive extremes representing increases in carbon uptake, while negative extremes being indicative of loss of carbon uptake \citep{Zscheischler_GRL_2014}.
While extremes in climate have been extensively studied \citep{Reichstein_2013_climate_ext, Sonia_2020_CMIP5-6, Ban_PrcpExt_2015, Frank_ClimateExtremes_CC_2015, Flach_Climate_extreme_GPP_2020}, few studies have focused on extremes in GPP and their underlying drivers\citep{Zscheischler_GRL_2014, Xu_CC_extremes_2019, Flach_Climate_extreme_GPP_2020}.
In this study, we identify extremes in global carbon uptake by computing percentile-based thresholds \citep{Seneviratne_IPCC_2012}.
Described in detail in Section~\ref{sec:methods_extremes}, a positive (or negative) extreme in GPP is defined as anomalies in GPP that are greater (or less) than a selected percentile-based threshold of GPP anomalies.
The thresholds were computed by selecting 1\textsuperscript{st} and 99\textsuperscript{th} percentile anomalies in GPP, calculated at every land grid cell for each consecutive 25~years from the year 1850 through 2300.
The GPP extreme events are then defined as values of GPP anomalies above \textit{q} (positive extremes) or below \textit{$-$q} (negative extremes).
The schematic Figure~\ref{f:schematic_methods} illustrates the steps to compute GPP extremes.


Extremes in GPP could also be categorized as isolated vs temporally continuous extreme events.
The isolated and temporally continuous extremes in GPP are analogous to hot days and heatwaves.
Extreme events in GPP that are continuous in time represent a significant cumulative impact on carbon uptake.
Akin to the definition of temporally continuous heatwaves by \cite{Baldwin_ContHW_2019}, we define a temporally continuous extreme (TCE) event (Figure~\ref{f:tce_schematic}) in GPP as 
        (a) three or longer months (equal to a season length) of GPP extreme anomaly occurring consecutively, 
		(b) and are considered as single continuous event through any gaps of two months or shorter in duration (assuming that ecological
				recovery is unlikely in less than a season) beyond three months.
A GPP TCE event that occur after a season, i.e., three months or more, is considered a separated TCE.
Section~\ref{sec:methods_attribution} illustrates the method for attribution of GPP TCEs to individual and compound effects of climate drivers.

\subsection{Identification and Detection of GPP Extreme Events} 
\label{sec:methods_extremes}

\begin{figure}
 \centering
 \includegraphics[width=0.95\columnwidth]{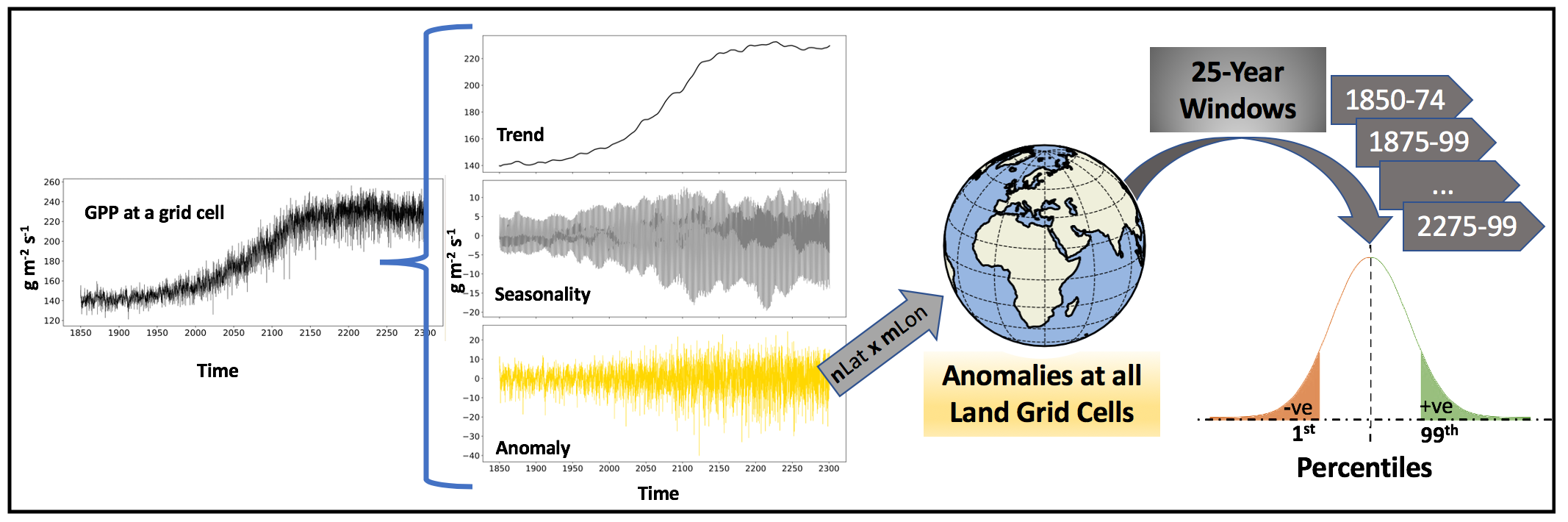}
 \caption{Schematic diagram for calculating thresholds in gross primary production (GPP). The anomalies are calculated at every grid cell by subtracting the nonlinear trend and modulated annual cycle from the GPP time series. The anomalies of every land grid cell ($nLat \times nLon$) for consecutive 25~year time windows were chosen to calculate probability distribution function of GPP anomalies. The 1\textsuperscript{st} and 99\textsuperscript{th} percentile values represent the global GPP threshold values for negative and positive extremes in GPP. }

\label{f:schematic_methods}
\end{figure}

The time series of GPP at any grid cell consists of a trend, seasonality, and anomalies (Figure~\ref{f:schematic_methods}) components.
We used singular spectrum analysis (SSA), a non-parametric spectral estimation method based on embedding a time series in vector space \citep{Golyandina_Book_20010123}, to extract signals with specific frequencies.
The trend in GPP at any grid cell captures long term change in mean GPP, which is influenced by long term changes in climate drivers, atmospheric CO\textsubscript{2} concentration, and LULCC.
Since  El Ni\~no-Southern Oscillation (ENSO), and to a lesser extent other large-scale drivers of climate variability, enhance the variability in terrestrial photosynthesis, the nonlinear trend at each grid cell was calculated by adding all the frequencies of 10 years and longer.
Hence, carbon cycle extremes in our study include the GPP anomalies induced by ENSO which is one of the largest modes of climate variability and peaks about every three to seven years \citep{Chylek_2018_ENSO} and exerts strong regional effects on the terrestrial carbon cycle \citep{Poveda_2001_ENSO}. 
The seasonality in GPP follows a periodic cycle of 12 months.
The conventional way to compute the annual cycle is to determine the mean climatology over a period, 
however, the climatology does not reflect the intrinsic nonlinearity of the climate-carbon feedback, especially under external forcing \citep{Wu_MAC_2008} because of the increased modulation of the annual cycle in GPP under business-as-usual rising CO\textsubscript{2} emissions.
We calculated the modulated annual cycle, which allows the annual cycle to change from year to year and consists of signals with a frequency of 12 months and its harmonics.
The anomalies at each grid cell were calculated by subtracting the trend and modulated annual cycle from the GPP time series (Figure~\ref{f:schematic_methods}).
Hence, GPP anomalies comprised of the high-frequency signals (\textless12~months) and the interannual variability (\textgreater12~months and \textless10~years).
We examined the probability distribution function (PDF) of GPP anomalies for all land grid cells for every 25-year time window from 1850 through 2300.
While the left tail of the PDF of GPP anomalies represents large losses in carbon uptake, the right tail portrays large gains in carbon uptake.
We chose the 1\textsuperscript{st} and 99\textsuperscript{th} percentile of all GPP anomalies as thresholds in the 25-year time windows to identify large losses and gains in carbon uptake.


The percentile-based thresholds (1\textsuperscript{st} and 99\textsuperscript{th}) for every 25-year time window yield the negative ($Th-$) and positive ($Th+$) threshold trajectories of GPP anomalies.
The extreme anomalies in GPP range from 425~GgC/month for the period 1850--74 to 840~GgC/month for 2275--99, as shown in Figure~\ref{f:threshold_all}.
Since GPP anomalies are a subset of GPP, increasing CO\textsubscript{2} fertilization, water use efficiency, and lengthening of growing season lead to increases in GPP (Figure~\ref{f:co2_traj}(c)) and the thresholds of GPP anomalies.
The simulations \emph{with} and \emph{without LULCC}, show higher magnitudes of $Th-$ than $Th+$, indicating that negative anomalies in GPP are stronger than positive for the same percentile.
Also, the values of the thresholds  for simulation \emph{with LULCC} were higher than \emph{without LULCC}.
To enable the comparison of simulations \emph{with} and \emph{without LULCC}, we selected one threshold trajectory, \emph{without LULCC} $TH+$ ($q$), to apply for calculations of all positive and negative extremes in the current study.
Thus, the positive GPP extremes are defined as GPP anomalies greater than $q$ and negative GPP extremes are the GPP anomalies less than $-q$ for both simulations, \emph{with} and \emph{without LULCC}.

Integral of negative (or positive) GPP extremes over land grid cells represents the total global loss (or gain) in carbon uptake per month.
The time series of frequency and extent (area) under GPP extremes was computed by integrating the count and area of grid cells under GPP extremes.

\subsection{Attribution to Climate Drivers}
\label{sec:methods_attribution}

Human activities, through fossil fuel emissions and land cover changes, modify the climate and climate-carbon feedbacks.
To attribute significant changes in carbon uptake and GPP to climate drivers, we computed linear regression of temporally continuous GPP extremes with anomalies in atmospheric precipitation ($Prcp$, composed of atmospheric rain and snow), precipitation minus evapotranspiration ($P-ET$), soil moisture ($Soilmoist$, up to 1-m depth), monthly maximum daily temperature ($T_\mathrm{max}$), monthly minimum daily temperature ($T_\mathrm{min}$), monthly mean daily temperature ($T_\mathrm{sa}$), and fire ($Fire$, total column level carbon loss due to fire) (Table~\ref{t:drivers_table}).

\begin{table}[]
\centering
\caption{Climate Drivers Considered for Attribution to GPP Extremes}
\label{t:drivers_table}
\begin{tabular}{lll}
\hline

\toprule
Symbol & Units & Description \\
\midrule
$Prcp$ & mm~s$^{-1}$ & Atmospheric rain + snow \\
$P$$-$$ET$ & mm~s$^{-1}$ & Precipitation minus Evapotranspiration \\
$Soilmoist$ & mm & Soil moisture in top 1-m depth \\
$T_\mathrm{max}$ & K & Monthly maximum daily temperature \\
$T_\mathrm{sa}$ & K & Monthly mean daily temperature \\
$T_\mathrm{min}$ & K & Monthly minimum daily temperature \\
$Fire$ & gC~m$^{-2}$~s$^{-1}$ & Total column level carbon loss due to fire \\
\bottomrule
\end{tabular}
\end{table}

The impact of climate drivers on GPP often has a lagged response because the terrestrial ecosystem has ingrained plasticity to buffer and push back effects of climate change \citep{Zhang_Lag_effects_2014}.
The controls of different climate drivers on GPP and its extremes is also dependent on location, timing, soil type and moisture, and vulnerability of land cover type \citep{Frank_ClimateExtremes_CC_2015}.
Therefore, at every location of GPP TCEs, we used linear regression to compute the correlation of TCEs with the cumulative lagged response of anomalies of every climate driver at time-lags of 0 to 12 months, for every 25 year time window from 1850 to 2300. 
However, lagged responses beyond four months were insignificant and thus we only report on one, two, and three lags.
Anomalies from past months were included in the computation of lagged correlations.
For instance, to compute a climate-carbon lag response of 3 months, the GPP extreme anomalies at month $t$ were correlated with the average of climate driver anomalies at $t-3$, $t-2$, and $t-1$ months.
We also investigated the GPP TCEs driven by triggers (climate drivers during onset of TCEs) and 
persistent climate drivers (considering entire duration of TCEs).
We identified the most dominant climate driver at any grid cell under simulations \emph{with} and \emph{without LULCC}, as the climate driver with the highest absolute correlation coefficient (significance value, $p < 0.05$).
The percent global distributions of dominant drivers for every time window were computed to inspect the changing patterns of dominant drivers at various lags and over time.

A GPP TCE event could be driven by one or a combination of climate drivers.
The study of compound events consisting of multiple different climate drivers leading to a GPP extreme improves our understanding of interactive effects of climate drivers \citep{Zscheischler_Compound_2018}.
While the climate extremes and carbon extremes often do not occur concurrently, the compound effects of climate drivers that are not climate extremes can nevertheless impact the carbon cycle \citep{Pan_2020_CarbonClimateExtremes}.
We studied the compound effects of water ($Prcp$, $P-ET$, and $Soilmoist$)-, temperature ($T_\mathrm{max}$, $T_\mathrm{sa}$, and $T_\mathrm{min}$)- and fire ($Fire$)-driven TCEs in GPP.

\section{Results}
\subsection{Detection and Identification of GPP Extremes} 
\label{sec:results_extremes}

\begin{figure}
 \centering
 \includegraphics[trim = {0.1cm 0.1cm 0.1cm .95cm},clip,width=0.95\columnwidth]{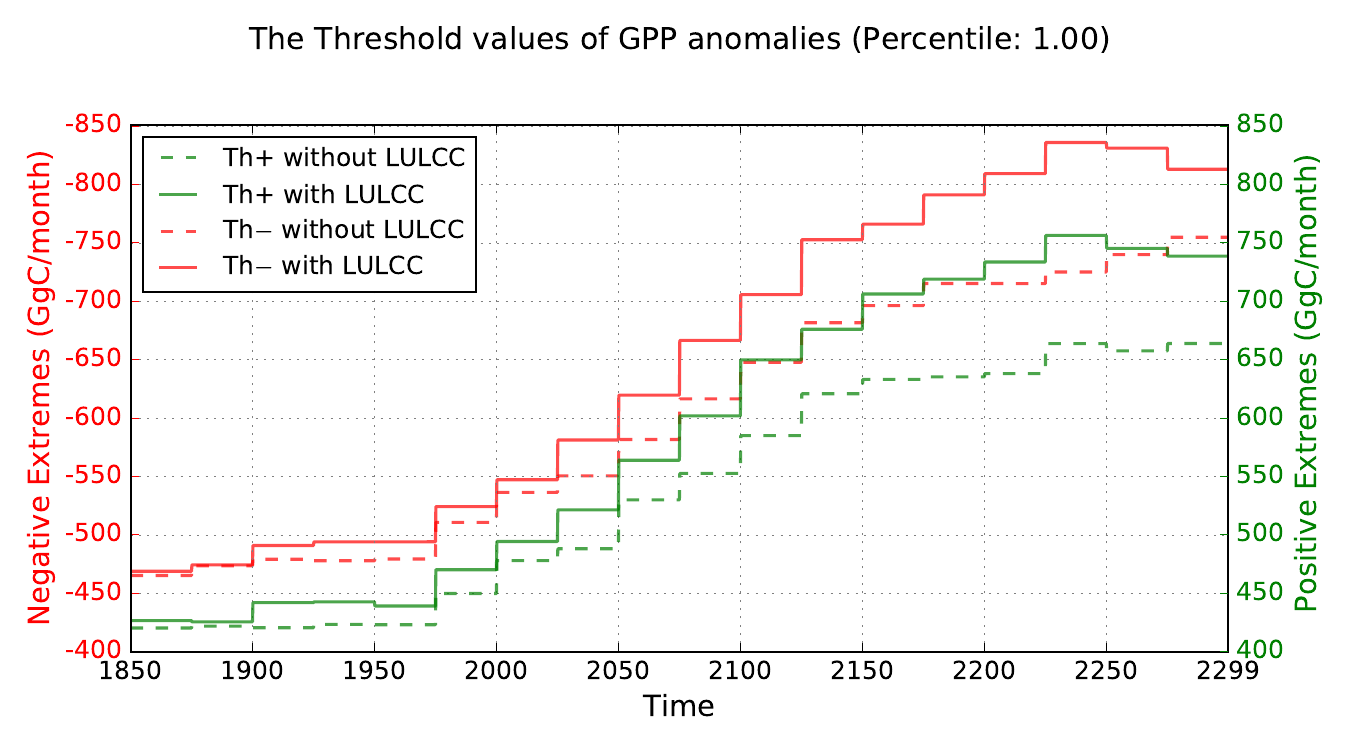}
 \caption{Thresholds of GPP extreme events \emph{with} and \emph{without LULCC} from 1850--2299. 
 The figure shows increasing thresholds of negative and positive GPP 
 extreme events based on 1\textsuperscript{st} and 99\textsuperscript{th} percentile, respectively. 
 The percentiles are calculated for global GPP anomalies for every time window (of 25~years) from 1850--2299. The red color represents negative thresholds or Th$-$ and green represents Th+. The solid and dashed lines represent the simulations \emph{with} and \emph{without LULCC} respectively.}

\label{f:threshold_all}
\end{figure}

The 1\textsuperscript{st} and 99\textsuperscript{th} percentiles of the PDF of GPP anomalies for all land grid cells in 25-year time windows were used to compute the threshold trajectories for\emph{with} and \emph{without LULCC} simulations.
Figure~\ref{f:threshold_all} shows trajectories of positive and negative GPP extremes for simulations \emph{with} and \emph{without LULCC}. 
The rising atmospheric CO\textsubscript{2} concentrations and increasing CO\textsubscript{2} fertilization, water use efficiency and lengthening of growing seasons \citep{Bonan_Book_2015, Lawrence_2012_lulcc_cesm1bgc} lead to an increase in the global GPP and GPP anomalies.
Consequently, $Th+$ and $Th-$ increased in both simulations; for the simulation \emph{without LULCC} $Th+$ increased from 420~GgC/month during 1850--1874 to 664~GgC/month during 2275--2299; the corresponding values for $Th+$ \emph{with LULCC} were 426 and 739~GgC/month, for $Th-$ \emph{without LULCC} were $-$465 and $-$755~GgC/month, and for $Th-$ \emph{with LULCC} were $-$469 and $-$813~GgC/month.
The increasing magnitude of thresholds for GPP extremes over time highlights the intensification of GPP extremes over time.
For the same percentile, the magnitude of negative thresholds were larger compared to positive thresholds.
Hence, negative anomalies in GPP or losses in carbon uptake were much
larger than gains in carbon uptake, primarily due to negative impact of increasing anthropogenic CO\textsubscript{2} and LULCC on carbon cycle anomalies.
Higher thresholds for the simulation \emph{with LULCC} despite lower GPP (Figure \ref{f:co2_traj}(c)) compared to the simulation \emph{without LULCC}, highlights that increasing magnitude of GPP anomalies were likely due to LULCC and wood harvest.
Higher anomalies GPP is potentially due to large reductions in the area of tree PFTs of primary vegetation (forests) in the RCP~8.5 LULCC scenario, $-3.5 \times 10^6$~km\textsuperscript{2} from 1850 to 2100.
The decrease in primary vegetation is associated with large increases in crop and grass areas, leading to a reduction of ecosystem carbon of $-$49~PgC from 1850 to 2100.
The global wood harvest carbon flux increased to 4.2~PgC\,year\textsuperscript{-1} \citep{Lawrence_2012_lulcc_cesm1bgc} in the year 2100, and then was kept at a constant harvest rate from 2100 to 2300.
The legacy effects of human land cover change and continued wood harvest becomes more visible beyond 2100 when the difference in annual GPP of both simulations widens (Figure \ref{f:co2_traj}(c)).
As a result of enhanced variability in GPP \emph{with LULCC}, we saw a larger magnitude of negative and positive thresholds for the simulation \emph{with LULCC} (Figure~\ref{f:threshold_all}).

\begin{figure}
 \subfloat[Intensity of Global extremes \emph{without LULCC} from 1850--2299]{\includegraphics[trim = {0.1cm 0.1cm .1cm .95cm},clip,width=0.95\columnwidth]{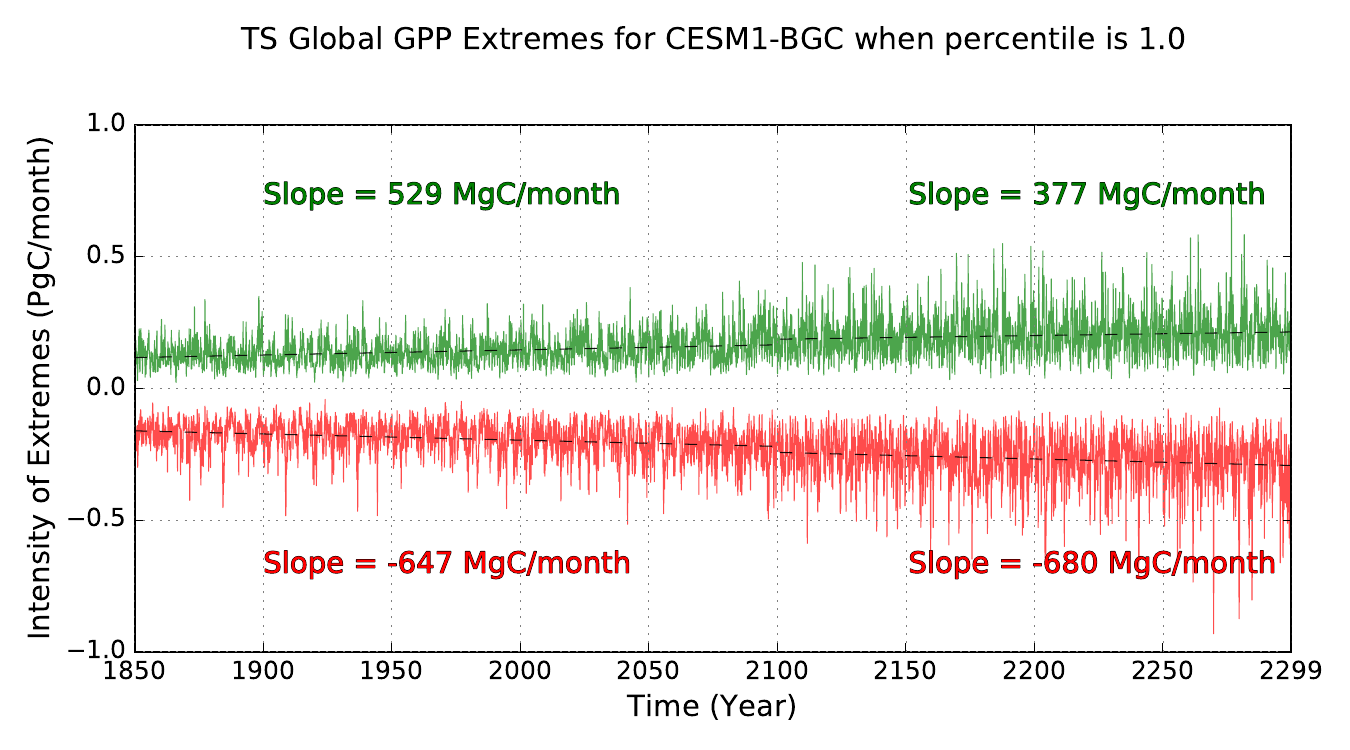}}\label{f:intensity_wo_lulcc_rel_pos_wo_lulcc}
 
 \subfloat[Intensity of Global extremes \emph{with LULCC} from 1850--2299]{\includegraphics[trim = {0.1cm 0.1cm .1cm .95cm},clip,width=0.95\columnwidth]{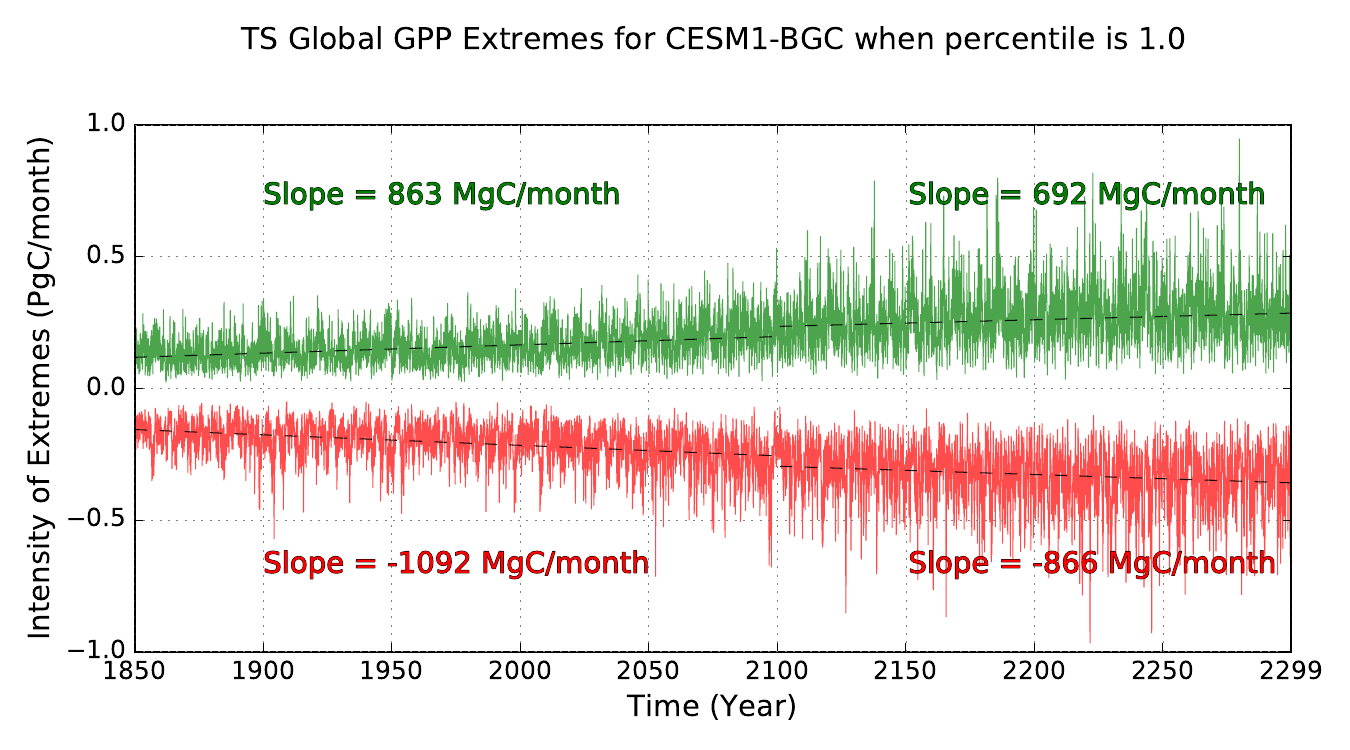}}\label{f:intensity_w_lulcc_rel_pos_wo_lulcc}
 
 \caption{Monthly time series of intensity of global GPP extreme events for the simulation \emph{without LULCC} (a) and \emph{with LULCC} (b) from 1850--2299. The positive GPP extremes, GPP anomalies $>$ $q$, are represented in green color and the negative extremes, GPP anomalies $<$ $-q$, are shown in red color. The rate of increase of positive GPP extremes \emph{without LULCC} are 529~MgC/month from 1850--2099 and 377~MgC/month from 2100--2299 (a). The corresponding rates for the growth of negative extremes are $-$647~MgC/month and $-$680~MgC (a). The rate of increase of positive GPP extremes \emph{with LULCC} are 863~MgC/month from 1850--2099 and 692 MgC/month from 2100--2299 (b). The corresponding rates for the growth of negative extremes are $-$1092~MgC/month and $-$866~MgC (b).}
 \label{f:intensity_rel_pos_wo_lulcc}
\end{figure}


The global time series of the intensity of losses and gains in carbon uptake, calculated by integrating negative and positive extremes in GPP, are shown in Figures~\ref{f:intensity_rel_pos_wo_lulcc}(a) and \ref{f:intensity_rel_pos_wo_lulcc}(b) for simulations \emph{without} and \emph{with LULCC}, respectively.
Compared to the simulation \emph{without LULCC}, the total additional loss of global carbon uptake due to LULCC was $-$46.53~PgC for period 1850--2100 and $-$141.76~PgC for 2101--2300.
The respective difference in total global GPP (\emph{with LULCC} minus \emph{without LULCC}) was $-$676~PgC for 1850--2100 and $-$1416~PgC for 2101--2300; and relative to the total global GPP, the additional losses in total carbon uptake due to LULCC increased from 6.9\% (1850--2100) to 10\% (2100--2300).
Hence, LULCC impacts global carbon cycle by reducing the total global GPP and increasing losses in carbon uptake during GPP extremes.
The rates of increase in the intensity of positive extremes in GPP for the simulation \emph{without LULCC} were 529 and 377~MgC/month for the periods 1850--2100 and 2101--2299, respectively, and for the simulation \emph{with LULCC} were 863 and 692~MgC/month, respectively.
The corresponding rates of increase in the intensity of negative extremes in GPP for the simulation \emph{without LULCC} were $-$647 and $-$680~MgC/month, respectively, and for the simulation \emph{with LULCC} were $-$1092 and $-$866~MgC/month, respectively.
The changes in the rates of the intensity of positive extremes in GPP for the simulation \emph{with LULCC} were analogous to the simulation \emph{without LULCC}.
However, the intensity of negative extremes in GPP for the simulation \emph{with LULCC} shows a decrease beyond 2100, possibly due to non-increment of wood harvest rate and a constant PFT distribution at the year 2100 values for the period from 2100 through 2300 which decreases the relative variability in GPP after 2100.
The larger intensity of negative extremes in GPP for the simulation \emph{with LULCC} than \emph{without LULCC} could result from increased wood harvest and land conversions to agriculture, pastures, and urban areas.
The rate of increase in the intensity of positive extremes in GPP was higher in the simulation \emph{with LULCC} than \emph{without LULCC} probably due to large re-growth of secondary forests in the regions of the Eastern U.S., Europe, Africa, and South America \citep{Hurtt_2011_Clim_Change_LULCC}.
In the simulation \emph{with LULCC}, by the year 2100 under RCP~8.5, the LULCC transitions resulted in high-density croplands in the U.S., Europe, and South East Asia; high-density pastures in the Western U.S., Eurasia, South Africa, and Australia \citep{Hurtt_2011_Clim_Change_LULCC}.
Primary forests were present in high northern latitudes and parts of Amazonia.
Hence, LULCC forcing coupled with CO\textsubscript{2} forcing under RCP~8.5 and ECP~8.5 intensified both losses and gains in carbon uptake during extremes in GPP, with net losses in carbon uptake dominating the net climate-carbon response.

\begin{figure}
 \centering
 \includegraphics[trim = {0.1cm 0.1cm 0.1cm .25cm},clip,width=0.85\columnwidth]{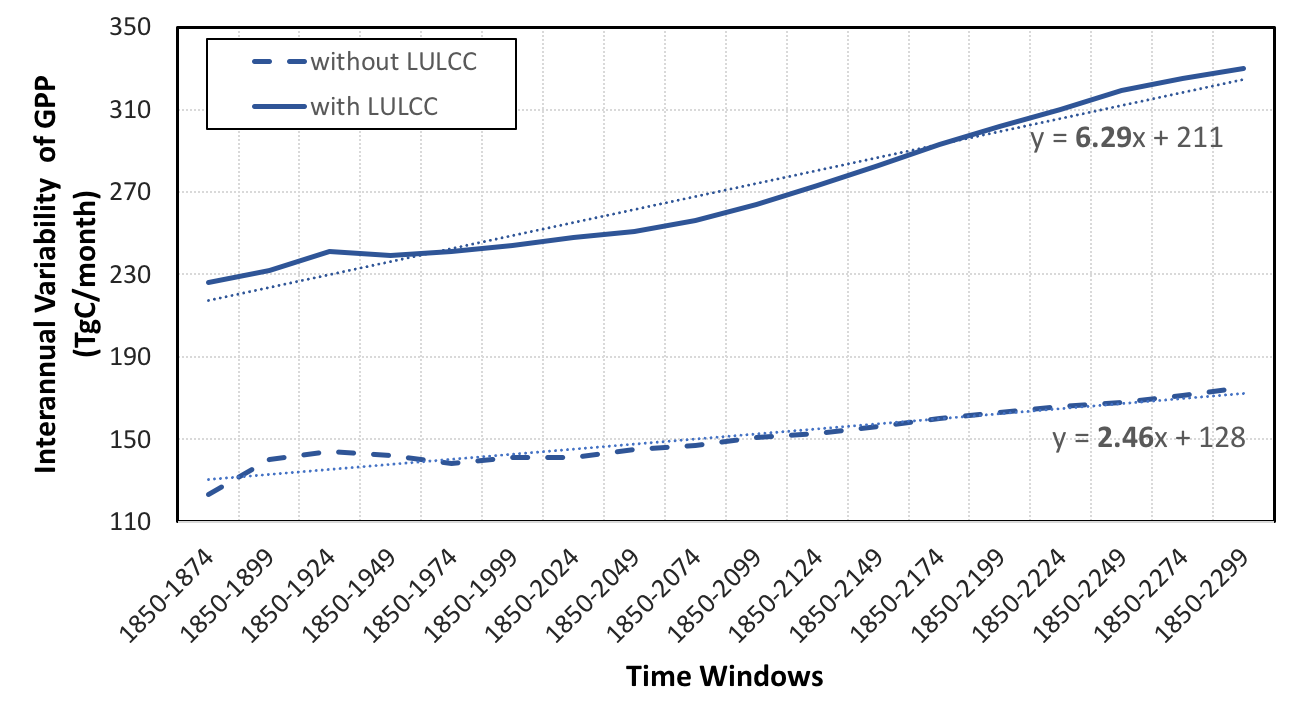}
 \caption{Global interannual variability (IAV) of GPP \emph{with} and \emph{without LULCC}. 
 The unit of IAV is $10^{12}$~gC. 
 The IAV is calculated from 1850 as the base year to 25~year increments, as shown on the $x$-axis. 
 The solid line represents the IAV of GPP for \emph{with LULCC} and dashed line represents the IAV of GPP for \emph{without LULCC}. 
 The linear fits represented by dotted lines show rates of increase of IAV of GPP \emph{with LULCC}, 
 which is higher by a factor of 2.5 compared to the simulation \emph{without LULCC}.}
\label{f:global_gpp_IAV}
\end{figure}

The rate of increase of intensities of positive and negative extremes in GPP was stronger for the simulation \emph{with LULCC}, although the total GPP was less than the simulation \emph{without LULCC} (Figure~\ref{f:co2_traj}(c)).
Stronger intensities of GPP extremes \emph{with LULCC} were due to the larger interannual variability (IAV) in GPP \emph{with LULCC} compared to \emph{without LULCC} (Figure~\ref{f:global_gpp_IAV}).
LULCC alters the climate by modifying the biogeochemical and biogeophysical processes, which further affects the carbon cycle.
For example, in semiarid climates, loss of vegetation cover increases the surface albedo, which increases the reflected solar radiation and cools the surface climate that weakens the boundary layer, reducing the probability of precipitation that creates the dry climate and reduces plant productivity.
Biogeochemical processes include uptake of carbon during photosynthesis at an increased atmospheric concentration of CO\textsubscript{2} and loss of carbon during respiration in a warmer climate.
Biogeophysical and biogeochemical processes do not occur in isolation and depend on the hydrologic cycle \citep{Bonan_Book_2015}.
With increases in atmospheric CO\textsubscript{2} concentration, the climate becomes warmer and increases the intensity of precipitation and accompanying precipitation extremes \citep{Ban_PrcpExt_2015,Gorman_Prcp_2015}.
Clearing of land and deforestation has led to and will lead to cooling in high latitudes, warming in tropics, and uncertain changes in mid-latitudes \citep{Lawrence_LUMP_2016}.
As a result, human LULCC increases the regional heterogeneity in vegetation that alters the climate drivers \citep{Ichii_IAVgpp_2005}, further increasing the global interannual variability and anomalies of GPP as well as the spatially heterogeneous distribution of GPP extremes.
Therefore, the effect of both CO\textsubscript{2} and LULCC forcing, represented in \emph{with LULCC}, increases the variability of biogeochemical and biogeophysical feedbacks, thus resulting in increased IAV of GPP in \emph{with LULCC} than \emph{without LULCC}.

The regional atmospheric circulation and climate change lead to spatial variations in the distribution of GPP extremes and the intensities of GPP extremes.
Since the definition of GPP extremes is based on the threshold trajectory of 1\textsuperscript{st} percentile global anomalies of GPP for consecutive 25-year time windows in the simulation \emph{without LULCC} (Figure~\ref{f:threshold_all}), the total number of grid cells under positive GPP extremes were constant at around 64,000 (for every 25-year time windows) while they vary for all other scenarios.
The total number of grid cells and area affected by negative extremes in
GPP for the simulation \emph{with LULCC} were largest, possibly because of increased negative feedback of climate variability on the carbon cycle due to the cumulative CO\textsubscript{2} and LULCC forcing.
Relative to the frequency of positive extremes in GPP for the year 1850 (\emph{without LULCC}), the frequency of positive extremes (\emph{with LULCC}) increased by 17\% and 28\% for the periods 1850--2100 and 2101--2300 respectively; and the respective growth rates of negative extremes (\emph{with LULCC}) were 13\% and 19\%.
For \emph{with LULCC}, growth rates for the area affected by positive GPP extremes were 16\% and 28\%, and for negative GPP extremes at 12\% and 20\% during 1850--2100 and 2101--2300 respectively. 
Higher growth rates of frequency and area-affected by positive extremes in GPP were probably due to increases in secondary forest cover; however the losses in the expected carbon uptake due to LULCC were larger than gains.

\subsection{Attribution to Climate Drivers}  
\label{sec:results_attribution}


\begin{figure}
 \subfloat[The mean duration of TCE events from 1850--2299]{\includegraphics[trim = {0.1cm 0.1cm .1cm .95cm},clip,width=0.95\columnwidth]{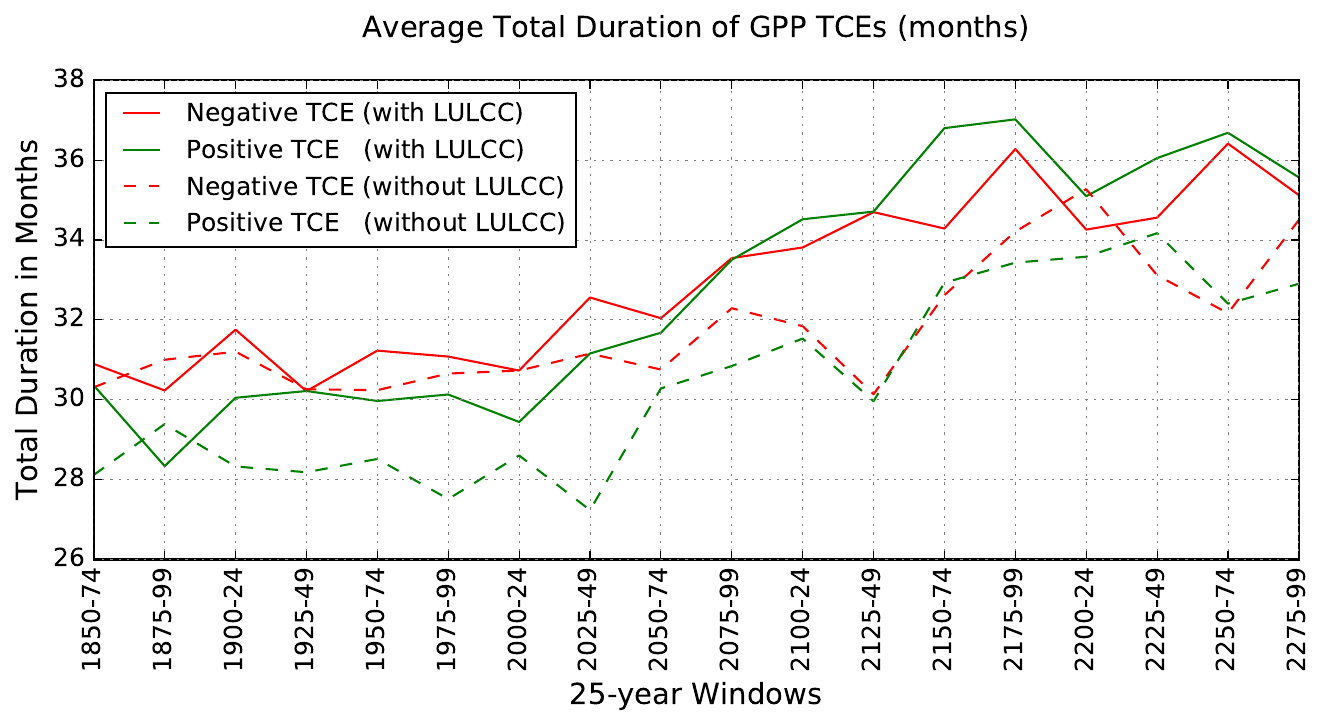}}\label{f:ts_tce_len_mean}
 
 \subfloat[Density of TCE events for a 25-year period per century]{\includegraphics[trim = {0.1cm 0.1cm .1cm .95cm},clip,width=0.95\columnwidth]{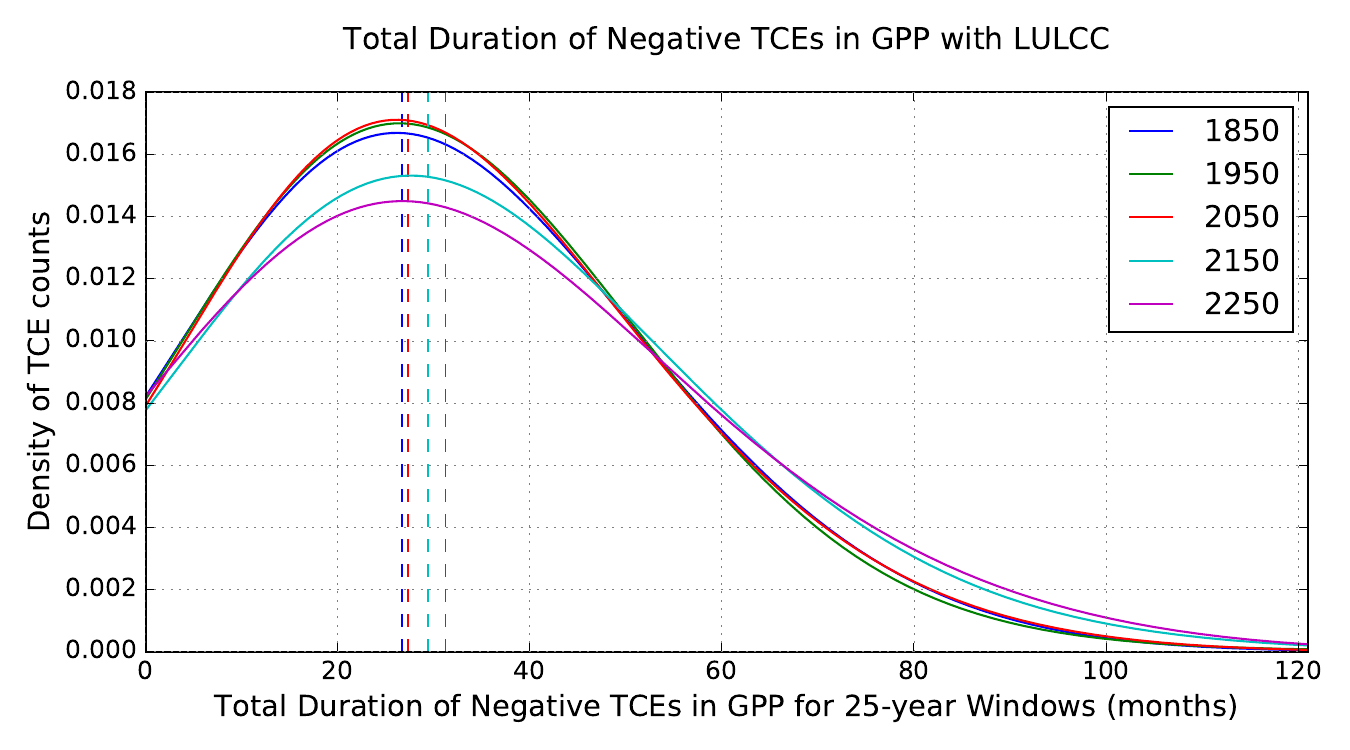}}\label{f:tce_kde_neg_len}
 
 \caption{The mean duration of positive (shown in green) and negative (shown in red) TCEs for both the simulations, \emph{with} (solid lines) and \emph{without LULCC} (dashed lines), for 25-year periods (a). The probability density of counts of total number of months under a negative TCEs in 25 years or 300 months (as shown in $x$-axis) for 25-year windows starting at the years 1850, 1950, 2050, 2150 and 2250 \emph{with LULCC} (b). The dashed vertical lines shows the shifting of mean duration of negative TCEs to right, highlighting that the TCEs are getting longer over time.}
 \label{f:tce_stats}
\end{figure}

The attribution of climate-driven extremes in GPP were performed for GPP TCE events, which are time-continuous GPP extremes meeting the criteria described in section~\ref{sec:methods_begin}.
Variability in terrestrial carbon cycle intensified the climate-driven GPP TCEs under the combined forcing of human-induced LULCC and anthropogenic CO\textsubscript{2} emissions.
The mean duration (Figure~\ref{f:tce_stats}(a)) and standard deviation (Figure~\ref{f:ts_tce_len_std}) of negative and positive TCEs in GPP for the simulation \emph{with LULCC} were greater than the simulation \emph{without LULCC}.
In addition, the duration of TCEs lengthened over time, with more long-duration TCEs and fewer short-duration TCEs as time progressed.
Figure~\ref{f:tce_stats}(b) shows the density plot of the count of negative TCEs in GPP vs.\ the total duration of GPP TCEs in 25-year windows for the simulation \emph{with LULCC}.
The increasing mean duration of the negative TCEs in GPP in the future likely causes a larger reduction in the carbon uptake, which has significant implications for carbon storage and the carbon budget, and it can negate the positive feedback of increasing CO\textsubscript{2} fertilization under RCP~8.5 and ECP~8.5 CO\textsubscript{2} scenarios.

\begin{figure}
 \centering
 \includegraphics[trim = {0.1cm 0.1cm .1cm .65cm},clip,width=0.95\columnwidth]{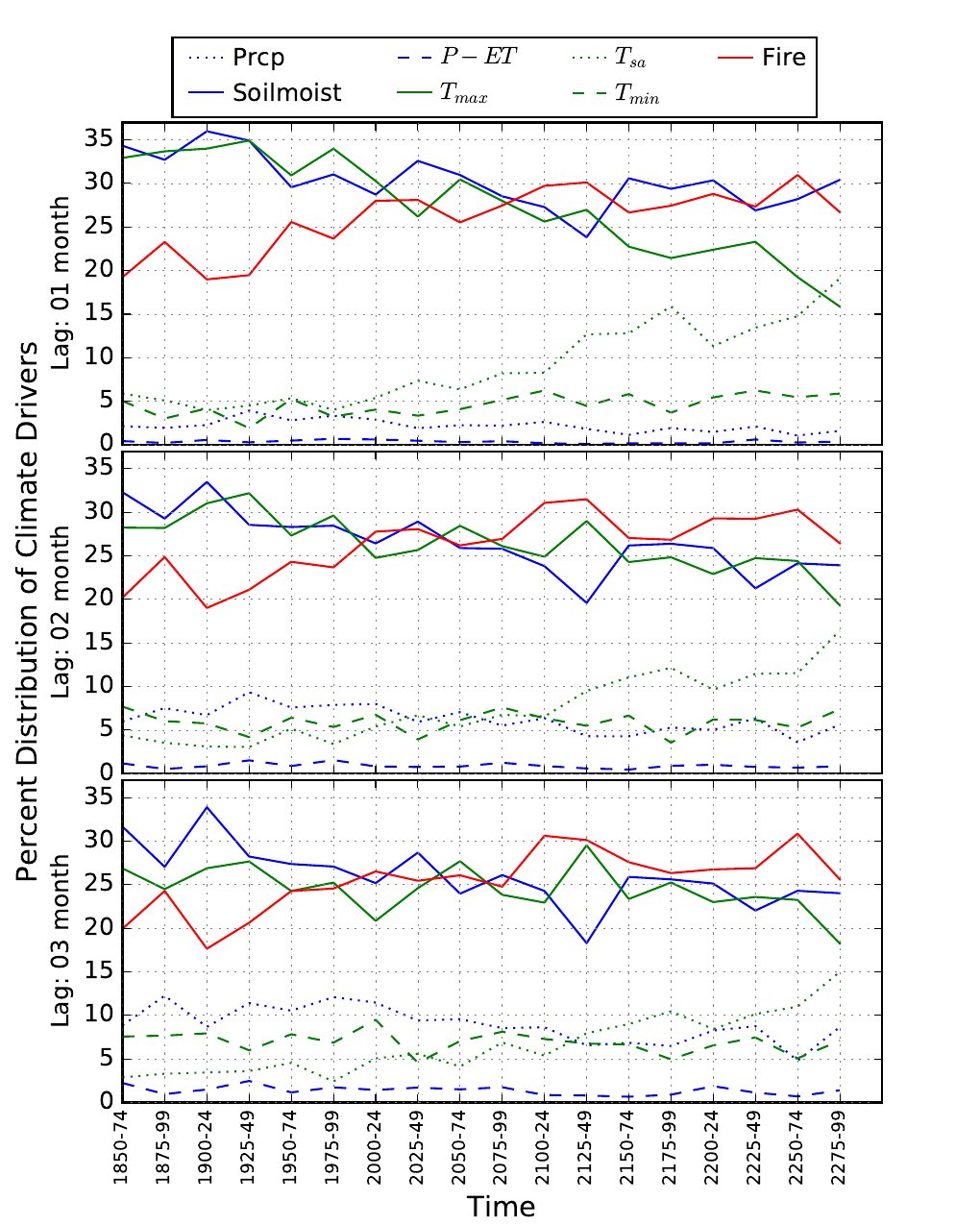}
 \caption{Percent distribution of global dominant climate drivers \emph{with LULCC} for every time window from 1850--2299. 
 For a particular lag month (1, 2, 3, etc.), a climate driver with highest correlation coefficient ($p$\,$<$\,$0.05$) with carbon cycle TCEs at any grid cell is called a dominant climate driver.}
\label{f:per_dom_w_lulcc}
\end{figure}
 
Extremes in carbon cycle and climate extremes often do not occur simultaneously \citep{Pan_2020_CarbonClimateExtremes}.
In this study, we first detect temporally continuous extremes (TCEs) in GPP and then attribute the climate drivers using linear regression.
Pearson's correlation coefficients and corresponding significance values ($p$-values) were computed between every climate driver and extreme anomalies in GPP during TCEs.
The dominant climate driver at every land grid cell was defined by the maximum absolute correlation coefficient of climate drivers ($p < 0.05$).
To consider the response of the ecosystem to prevailing climatic conditions, linear regressions were performed at multiple lags between extreme anomalies in GPP and climate driver anomalies.
Figure ~\ref{f:per_dom_w_lulcc} (and Figure~\ref{f:per_dom_wo_lulcc}) show the percent distribution of dominant climate drivers for every 25-year time window for both simulations, \emph{with LULCC} and \emph{without LULCC}, respectively, at lags of 1, 2, and 3 months.
The relationship of water availability (soil moisture) with TCEs in GPP was globally dominant during most time windows (Figure~\ref{f:per_dom_w_lulcc}).
Reduction in soil moisture leads to an anomalous reduction in photosynthesis \citep{Frank_ClimateExtremes_CC_2015}.
Soil moisture is the most dominant driver of negative carbon cycle extremes, highlighting a strong positive correlation of plant productivity with soil moisture, as shown in Figure~\ref{f:rgb_neg_w_lulcc_dri_win6_lag1}(b).
Temperature ($T_\mathrm{max}$ and $T_\mathrm{sa}$) and $Fire$ have a negative correlation with TCEs in GPP (spatial distribution shown in Figures~\ref{f:rgb_neg_w_lulcc_dri_win6_lag1}(d) and \ref{f:rgb_neg_w_lulcc_dri_win6_lag1}(e)), where an anomalous increase in these drivers leads to loss in carbon uptake.  $T_\mathrm{max}$ was dominant at a lag of 2 months through 2100, but $Fire$ shows dominance at higher lags, especially beyond 2100 in the simulation \emph{without LULCC}.
The percent count of dominant temperature drivers, $T_\mathrm{max}$ and $T_\mathrm{sa}$, increased over time, especially after 2100.
Hence, carbon extremes driven by hot conditions will have the largest increase in a warming climate, especially after 2100.
The lagged response of anomalously hot air temperatures and lack of soil moisture creates hot and dry conditions suitable for fire events.
Fire events driven by hot and dry conditions increase at a higher rate at all lags in the simulation \emph{with LULCC} than the simulation \emph{without LULCC}.
After 2100 in the simulation \emph{with LULCC}, fire stands out as the dominant driver at all lags, highlighting that the human-induced LULCC will make ecosystems more vulnerable to fire events.
The results of our study are consistent with the findings of \citet{Williams_TonGPP_2014} and \citet{Zscheischler_GRL_2014}, suggesting that the declines in GPP with warm temperature extremes are due to dependence of GPP on soil moisture, and to the strong negative correlation between soil moisture and temperature.
The rising temperature under global warming will create a warmer environment in the future, increasing the risk associated with heatwaves and their impacts on the ecosystem.
A decline in precipitation and soil moisture compounded with warm temperature may cause an unprecedented increase in loss of carbon uptake and potentially reduce the terrestrial carbon sink.

\begin{sidewaysfigure}
 \centering
 \includegraphics[width=0.95\columnwidth]{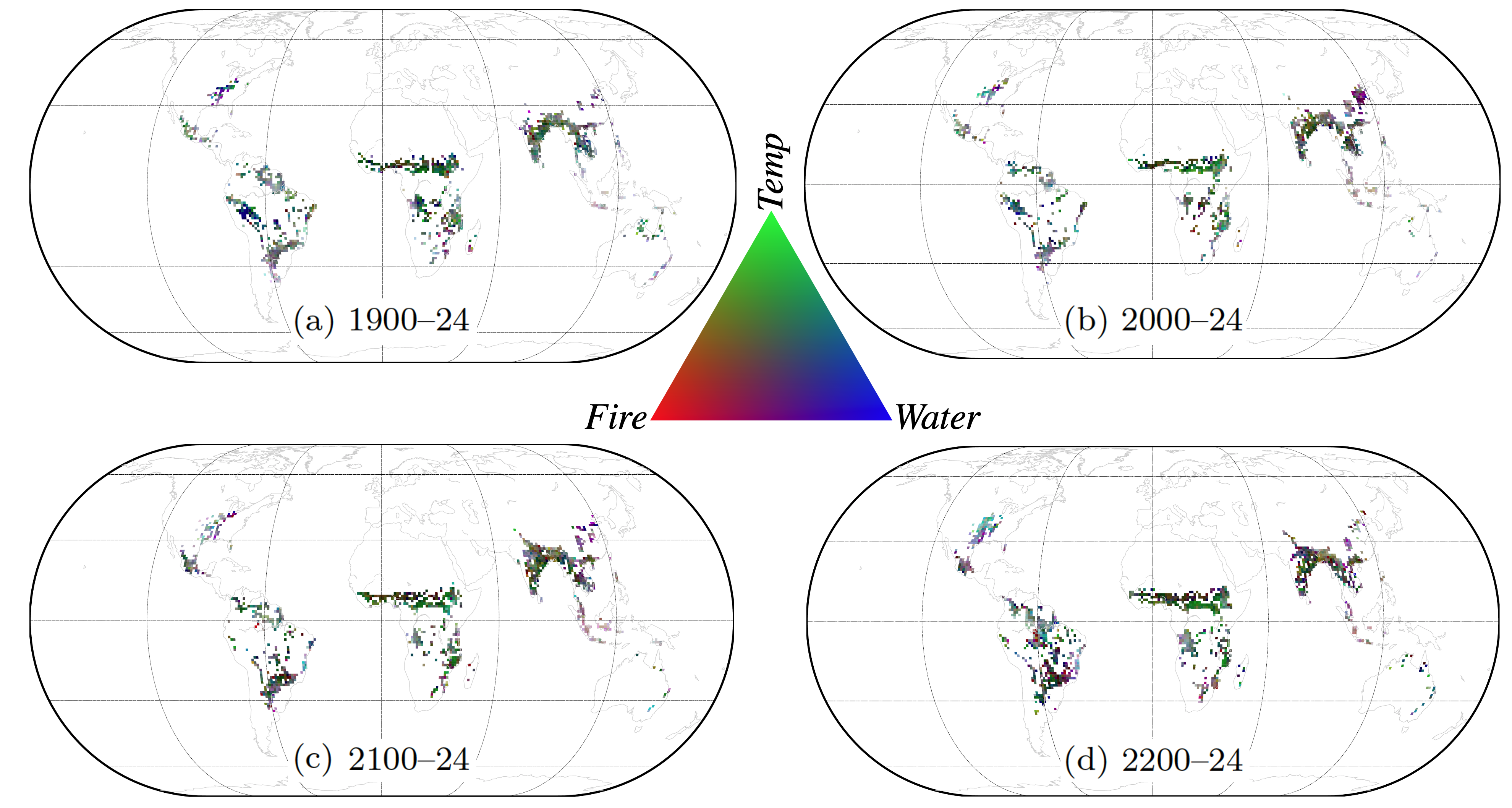}
 \caption{Spatial distribution of climate drivers causing negative TCEs in GPP for \emph{with LULCC} for four 25-year time windows, (a) 1900--24, (b) 2000--24, (c) 2100--24, and (d) 2200--24. The climate drivers are pooled in three colors, red, green, and blue. Red ($Fire$) is for loss of carbon due to fire, green ($Temp$) represents monthly maximum, mean, and minimum daily temperatures ($T_\mathrm{max}$, $T_\mathrm{sa}$, $T_\mathrm{min}$ respectively), Blue ($Water$) includes monthly means of soil moisture, precipitation and $P$$-$$E$ (precipitation minus evapotranspiration). The results shown here are at 1 month lag.}

\label{f:rgb_spatial_w_lulcc}
\end{sidewaysfigure}
\begin{figure}
 \centering
 \includegraphics[trim = {.08cm .1cm .1cm .95cm}, clip, width=0.85\columnwidth]{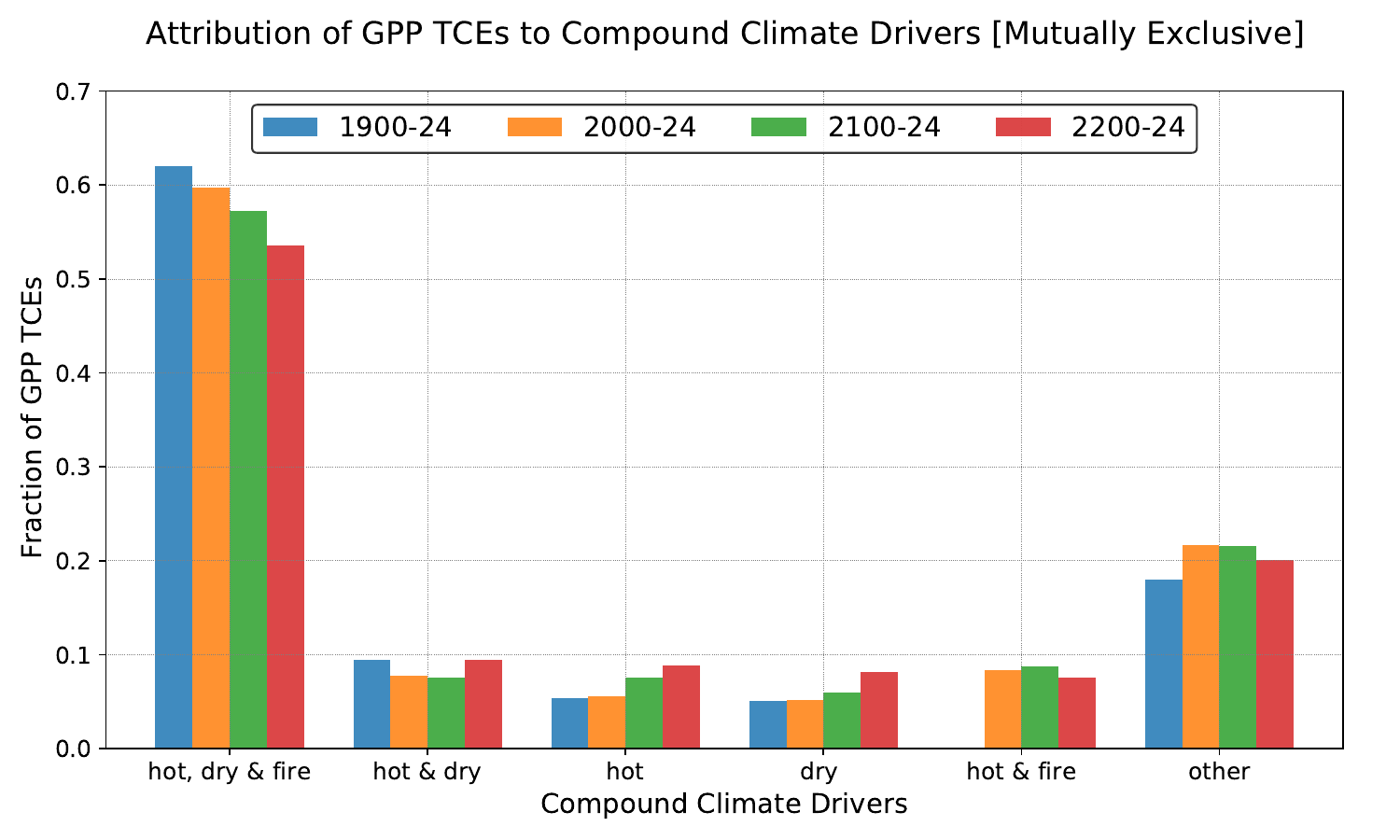}
 \caption{Attribution of temporally continuous extreme events in GPP to compound effect of climate drivers for \emph{with LULCC} at lag of 1~month for 25-year time windows, (a) 1900--24, (b) 2000--24, (c) 2100--24, and (d) 2200--24. The fractions are mutually exclusive, i.e., events driven by \textit{hot and dry} climate is not counted in either \textit{hot} or \textit{dry} climate driven events. Any location could be affected by one or compound climatic conditions. For example, a carbon cycle extreme could be driven by any combination of hot or cold, dry or wet, and with or without fire. We only show the combination of driving climate drivers that have total fraction of more than 0.05.}
\label{f:rgb_neg_w_lulcc_frac}
\end{figure}

The combined effect of climate drivers often has a larger impact on extremes in carbon cycle than the simple addition of individual climate drivers \citep{Zscheischler_Compound_2018,Zscheischler_GRL_2014, Frank_ClimateExtremes_CC_2015, Flach_Climate_extreme_GPP_2020}.
To capture the compound effect of climate drivers (Table~\ref{t:drivers_table}) on GPP TCE events, the climate drivers were grouped under fire, temperature, and water-driven extremes (Section~\ref{sec:methods_attribution}).
Figures~\ref{f:rgb_spatial_w_lulcc} and \ref{f:rgb_neg_wo_lulcc} show the spatial distribution of climate drivers at a lag of one month for simulations \emph{with} and \emph{without LULCC}, respectively.
We observed an increase in the number of TCEs in GPP and changes in the spatial distribution of GPP TCEs and climate drivers for time periods 1900--24, 2000--24, 2100--24, and 2200--24 in the simulation \emph{with LULCC} (Figure~\ref{f:rgb_spatial_w_lulcc}).
Most locations experienced TCE events due to the combined effects of all the climate drivers. 

We computed all combinations of water (dry or wet), temperature (hot or cold), and fire-driven climatic conditions that cause GPP TCEs.
For brevity, we report results for one 25-year time window one per century.
The fractions (larger than 0.05) of total negative TCEs in GPP driven by mutually exclusive climate conditions are shown in Figure~\ref{f:rgb_neg_w_lulcc_frac}.
More than half of the total TCEs in GPP occurred when the environmental conditions are hot and dry, resulting in fire events and account for large reductions in GPP (Figure~\ref{f:rgb_spatial_w_lulcc} and~\ref{f:rgb_neg_w_lulcc_frac}).
Similar to other studies \citep{Zscheischler_GRL_2014, Frank_ClimateExtremes_CC_2015}, a stronger correlation between warm and dry climate has a substantial impact on the terrestrial carbon cycle.
The increased fraction of TCEs attributed to hot climate over time is possibly due to combined attribution of $T_\mathrm{max}$ and $T_\mathrm{sa}$, as shown in Figure~\ref{f:per_dom_w_lulcc}.
Figure~\ref{f:rgb_neg_w_lulcc_frac_inclusive} represents the total fraction of TCEs in GPP driven by mutually inclusive climatic conditions.
As global warming increases the temperature of the planet, ecosystems will become more vulnerable to the hot and dry climate and at increasing risk of fire events (Figure~\ref{f:rgb_neg_w_lulcc_frac}).

Here we highlight the temporal changes in the distribution of carbon cycle extremes and their climate drivers for the simulation \emph{with LULCC}.
East Asia experiences a large number of TCEs during the 21\textsuperscript{st} century (Figure~\ref{f:rgb_spatial_w_lulcc}) driven by dry climate and fire, which however declines in the 22\textsuperscript{nd} due to an increase in precipitation and available soil moisture.	
The Savannas near Congo in Africa display an increased vulnerability of vegetation to fire and hot temperatures.
The impact of LULCC in Savannas is prominent after 2100 (Figures~\ref{f:rgb_neg_wo_lulcc}(d) and \ref{f:rgb_spatial_w_lulcc}(d)) when the pattern of dominant drivers change from temperature-driven (green color) to a compound effect of increase in dry, hot, and fire conditions (gray color).
The contiguous United States (CONUS) experiences more GPP TCEs, especially in the 23rd century (Figure~\ref{f:rgb_spatial_w_lulcc}(d)), driven by hot temperature and water limitation (highlighted in cyan color).
Indonesia and its neighboring islands show increasing GPP TCEs attributed to a combination of hot temperatures and fire.
The number of negative TCEs in South America are approximately the same for periods in Figures~\ref{f:rgb_spatial_w_lulcc}(a), \ref{f:rgb_spatial_w_lulcc}(b) and \ref{f:rgb_spatial_w_lulcc}(c) except for the period 2200--24 (Figure~\ref{f:rgb_spatial_w_lulcc}(d) when the extremes show a large increase.
During 2200--24, South America shows a drastic increase in negative TCEs that is attributed primarily to dry, hot, and fire conditions (represented by gray color).
Africa experiences an increase in the number of GPP TCEs attributed to hot climate, and, in the far future (2101--2300), witnesses an increased frequency of fire events.
Australia's east coast also exhibits an increase in GPP TCEs because of water limitation, dry climate, and they are accompanied by fires in the far future.
The South and Southeast Asia (composed of India, Myanmar, Thailand, and Cambodia) experiences an increase in TCEs in GPP over time.
These Asian regions where primary forest were converted to cropland \citep{Hurtt_2011_Clim_Change_LULCC}, saw an increase in fire-driven extreme events in GPP, highlighting a rising vulnerability to fire events due to LULCC.

\section{Discussion}
\label{s:discussions}

\subsection{Regional Analysis of Climate Change Impact on Climate Drivers and Negative Carbon Cycle Extremes}
\label{sec:discuss_diff}
Most climate models show an increase in the interannual variability of land--atmosphere CO\textsubscript{2} exchange over time \citep{Keenan_IAV_2012}.
The increasing atmospheric CO\textsubscript{2} concentration influences climate through its radiative effect (i.e., Greenhouse effect) and indirectly through physiological effect (reduced plant transpiration) \citep{Cao_CO2_2010}.
The increasing radiative effects cause changes in circulation, impacting precipitation, soil moisture, and global scale water cycle.
Increase in precipitation and CO\textsubscript{2} fertilization and reduction in stomatal conductance could lead to an increase in GPP; however, droughts, heatwaves and fires curtail plant photosynthesis \citep{Swann_2016_drought}.
Under climate change, while most regions experience an increase in mean GPP, some regions exhibit a decline.


Reduction in growth rate of expected terrestrial carbon uptake below carbon emissions could lead to weakening of carbon sink capacity (Figure~\ref{f:diff_gpp_and_negext_wo_lulcc}).
Increase in total GPP and IAV in GPP over time contribute to the increase in magnitude of negative carbon cycle extremes.  However, some regions experience weakening of negative carbon cycle extremes often due to decline in total GPP, and reduction in IAV of GPP and climate drivers (Figure~\ref{f:climate_drivers_disc}).
Driven by large scale circulation changes, increasing temperature and decreasing precipitation and soil moisture in Central America (CAM) lead to forest mortality, decline in GPP and weakening of negative carbon cycle extremes.
The decrease in precipitation in the Northern South American Tropics (NSA) and Central Amazon Basin (CAB) and increase in precipitation in the Southwest Amazon Basin (SAB) is due to plant physiologic response.
\citet{Langenbrunner_Amazon_prcp_2019} investigated this phenomenon in the regions of the Amazon and the Andes and found that under increased atmospheric CO\textsubscript{2}, stomatal conductance decreases, reducing water loss through transpiration, which decreases the convective activity and causes a reduction in rainfall over the Amazon Basin (CAB, NSA) and increases moisture advection by low-level westward jets towards the Andes, leading to increases in the rainfall over the Andes (SAB).
Reducing variability and increasing precipitation over SAB increase available soil moisture increasing regional GPP, however with reduced IAV, resulting in weakening of negative carbon cycle extremes.
Although CAB witnesses a decline in precipitation it demonstrates an increase in GPP likely due to slight reduction in soil moisture and large CO$_{2}$ fertilization effect which compensates for the reduction in soil moisture. 
Hence, in CAB both GPP and IAV in GPP increases over time causing strengthening of negative carbon cycle extremes.
These regional patterns of changes in GPP, negative carbon cycle extremes, and climate drivers are seen across the globe. 
For example, the regional changes in Northern Guinea, Southern Guinea, Central Guinea and Indonesia resemble the processes in CAM, SAB, CAB and CAM, respectively.
Additional feedbacks due to LULCC -- as tropical forests are replaced with crops or grasslands that transpire less than forests, further reducing rainfall over the Amazon -- causes hotter and drier climate with an increasing risk of fire and larger losses of carbon uptake (Figures~\ref{f:diff_gpp_and_negext_w_lulcc} and~\ref{f:diff_sm_wo_lulcc}).

Our robust attribution analysis based on successive time windows captures the changes in the anomalies of climate drivers and carbon cycle.  During 1900--24 and 2000--24, most of the negative TCEs in GPP that occurred in SAB (Figure~\ref{f:rgb_spatial_w_lulcc}), were due to water limitation (highlighted by the blue color in the RGB maps).
With increased water availability in SAB \citep{Langenbrunner_Amazon_prcp_2019}, the effects of heat associated with temperature increases are mitigated and the number of negative carbon cycle extremes reduces.
The decrease in precipitation in NSA led to a reduction in soil moisture, resulting in a hot and dry climate, thus making the vegetation prone to fire at high temperatures (the compound effect is highlighted by the gray color).
The reduced GPP in NSA over time further leads to a fewer number of negative carbon cycle extremes (Figure~\ref{f:rgb_spatial_w_lulcc}).  
Large IAV and anomalies in GPP in CAB lead to increased concentration of negative TCEs in GPP driven possibly due to combined effect of increased hot, dry, and fire events (shown in  gray color).

\subsection{Drivers and Triggers of Carbon Cycle Extremes}
\label{sec:discuss_tce}

A grid cell could experience any number of extreme events of any length during successive 25 year time windows from 1850 through 2300. 
Our methodology helps identify long temporally continuous extremes which represent large magnitude carbon extremes and have higher significant regression coefficient.
We use this methodology to identify the prevailing climatic conditions that act as triggers for an extreme and the conditions that cause an extreme to persist.
We demonstrate the findings from Chaco Province in Argentina where the major plant functional type is broadleaf deciduous tree (43\%).

Under the simulation \emph{without LULCC} and the time period 2000--24, the total number of positive TCE events in GPP were seven with a total duration of 53 months, which were greater than the total five negative TCEs in GPP with a total duration of 40 months (Table~\ref{t:tce_wo_lulcc_ag}).
The total gain in carbon uptake was 35.17~TgC, which was greater than the total loss in carbon uptake of $-$28.15~TgC, resulting in a net gain in carbon during TCE events.
Under the simulation \emph{with LULCC}, during the same period of 2000--24 the total number of negative GPP TCEs (6 events, 57 months) were greater than positive TCEs (5 events, 35 months).
Thus, the total loss in carbon uptake ($-$35.86~TgC) \emph{with LULCC} was higher than the gain in carbon uptake (23.87~TgC).
While this region had a net gain in carbon uptake of 7.02~TgC for the simulation \emph{without LULCC}, and it had a net loss in carbon uptake of $-$11.99~TgC for the simulation \emph{with LULCC}.
The increase in negative extremes in GPP with net losses in carbon uptake demonstrates the role of human-induced LULCC in negatively affecting carbon uptake capacity.

We performed a qualitative investigation of every negative and positive TCEs in GPP (Figure~\ref{f:tce_wo_lulcc_dri_ano_ag_combined}) using normalized time series of GPP, GPP anomalies and anomalies of major climate drivers ($Prcp$, $Soilmoist$, $T_\mathrm{max}$, and $Fire$).
Most positive TCEs in GPP were driven by increase in precipitation, followed by rise in soil moisture and decline in $T_\mathrm{max}$.
Most negative extremes in GPP were driven by decline in precipitation and rising $T_\mathrm{max}$, which cause high evapotranspiration and loss of soil moisture, eventually creating dry and hot conditions that cause fire events.
Some of the negative extremes in these regions were also found to be driven by hot and dry events without fire.

To identify the triggers of carbon cycle extremes, we define the onset period of GPP TCEs as first-quarter of every TCE. 
The regression  onset period of GPP TCEs and anomalies of climate drivers were computed with consideration of the lagged response of climate drivers on GPP TCEs.
The lagged soil moisture anomalies were highly correlated ($p < 0.01$)
	with entire duration of TCE events (persistence, Table~\ref{t:reg_wo_lulcc_win6_cum_lag}), and the lagged precipitation anomalies were highly correlated with the onset (trigger, Table~\ref{t:reg_wo_lulcc_win6_cum_lag_trig}) of TCE events in the Chaco Province.
	
For the simulation with CO$_{2}$ only forcing (i.e. \emph{without
		LULCC}) during time period 2000--24, the correlation
coefficient, for the whole duration of GPP TCEs, was highest for soil
moisture anomalies at a lag of 1~month.
The precipitation anomalies had high positive correlations with GPP TCEs (i.e., reduction in precipitation was correlated with a decrease in GPP) at lags of 3 and 4~months.
$T_\mathrm{max}$ and $Fire$ had large negative correlations with GPP
anomalies (i.e., increase in temperature correlates with a decrease in
		GPP or increase in carbon uptake loss) at a lag of 2~months.
Hence, for a negative GPP TCE event, the compound effect of a decline in
precipitation followed by increase in temperature caused a reduction of soil moisture, resulting in hot and dry conditions, increasing the probability of occurrence of fire and causing a long negative TCE event ($Prcp > T_\mathrm{max} >  Soilmoist > \mathrm{TCE}$).
For the same region under simulation \emph{with LULCC}, the dominant trigger was $T_\mathrm{max}$ (Table~\ref{t:reg_w_lulcc_win6_cum_lag_trig}) followed by precipitation.
Therefore, human-induced LULCC coupled with increasing
CO\textsubscript{2} levels tends to enhance the vulnerability and loss
in vegetation due to hot and non-dry climate conditions, resulting in more negative extremes in GPP.
The analysis illustrates the strength of our methodology in identifying the evolution of carbon cycle extremes for multiple conditions and at fine resolution.

\subsection{Limitations of CESM1(BGC)}
\label{sec:discuss_limitations}

The ILAMB benchmarking scores \citep{Nate_2018_ilamb} of the carbon fluxes and climate drivers indicate that CESM1(BGC) is among the best CMIP5 models (Table~\ref{t:table_ilamb}).
However, the Community Land Model version 4 (CLM4), the land model used
in the CESM1(BGC) \citep{Oleson_CLM4_2010}, had some limitations which
could potentially impact some of our findings.
The simulated GPP by CLM4 had a positive bias across the globe compared to FLUXNET eddy covariance tower estimates due to lack of colimitation of GPP to canopy scaling and parameterization of leaf photosynthesis kinetics \citep{Bonan_2011_ConopyProcesses}.
With improvements in the model parameterizations of radiative transfer, leaf photosynthesis and stomatal conductance, canopy scaling of leaf processes, inclusion of multilayer canopy model, and updated maximum rate of electron transfer parameters \citep{Bonan_2011_ConopyProcesses, Bonan_2012_MultiLayerCanopy}, the positive bias in GPP was reduced.
With inclusion of vertically resolved CN model in CLM4.5 \citep{Koven_2013_CN}, the positive bias in GPP was further improved and terrestrial carbon storage increased consistently with observations.
In the current study focused on the patterns of carbon cycle extremes, the positive bias of GPP was likely captured by trend and removed for calculation of GPP anomalies. 
However, any associated increase in the IAV of GPP corresponding to positive bias of GPP would increase the magnitude of GPP anomalies and both negative and positive GPP extremes.
Thus, there is a potential to overestimate the magnitude of both negative and positive GPP extremes but the relative comparison is insightful and most likely not affected by this
		limitation of CLM4.

CLM4 lacks representation of dynamic crop, thus the cooling effects in
irrigated lands, changes in the sensible and latent heat, and soil
carbon change are not well represented in CLM4.0 but were improved in the CLM5.0 \citep{Lombardozzi_2020_Ag_CESM2}.
The simulated climate-carbon feedbacks in CLM4 assume an instantaneous response of plant physiological processes to changes in temperature \citep{Lombardozzi_2015_Temp_Acclimation}.
With inclusion of temperature acclimation, though the ecosystem carbon storage pool grew (mostly in higher latitudes), the effect on photosynthesis of tropical regions was minimal. 
Since most of the detected extremes in the tropical forest and other high
biomass regions, the lack of representation of agriculture and
temperature acclimation on GPP are not expected to have significant
effect on the analysis of extremes.

\section{Conclusions}

Using the fully-coupled Earth system model, CESM1(BGC), we analyzed the
development of extreme events in GPP and attributed those carbon cycle
extremes to climate drivers from the year 1850 through 2300 for
simulations \emph{with} and \emph{without LULCC}.
While both simulations were forced with RCP~8.5 and ECP~8.5 atmospheric
CO\textsubscript{2} concentrations, only the simulation \emph{with
	LULCC} had additional forcing from human-induced land use and land cover change (LULCC).
The changes in land cover directly modify the biogeophysical and
biogeochemical feedbacks of terrestrial vegetation and indirectly
through the physiological responses of vegetation on climate drivers,
		leading to increase in interannual variability in GPP and higher magnitude of GPP anomalies.
Relative to the simulation \emph{without LULCC}, the simulation
\emph{with LULCC} exhibited increased variability in GPP and higher intensity, duration, extent, and frequency of extremes in GPP.
These characteristics were greater for negative extremes in GPP than positive extremes, implying larger than expected losses in carbon uptake than carbon gains.
Although the total GPP for the simulation \emph{with LULCC} was less
than the total GPP \emph{without LULCC}, the simulation \emph{with LULCC}
showed a sharper decline of carbon uptake of $-6.9\%$ in the near future
(1850--2100) and $-10\%$ in the far future (2100--2300) with respect to
total GPP \emph{without LULCC}.
The interactive effects of LULCC with the RCP~8.5 and ECP~8.5 CO\textsubscript{2} forcing amplify the weakening of the net carbon uptake, and a reduction of land carbon sink capacity, which could greatly affect the global carbon budget.

Increasing atmospheric CO\textsubscript{2} concentrations drive growth in vegetation photosynthesis or GPP due to carbon fertilization, and reduction in stomatal conductance.
Therefore, most places exhibited increased GPP and higher interannual variability in GPP.
Due to circulation changes, a few regions witness a decrease in precipitation, creating a drier climate that when supplemented with warmer temperatures lead to decline in GPP, such as the regions of the eastern Amazon and Central America. 
Reducing magnitudes of GPP over time and analogous decreases in the interannual variability of GPP produce weakening of negative carbon cycle or GPP extremes in the Amazon.
Moreover, the weakening of other negative carbon cycle extremes could be a result of less variable and benign climatic conditions for vegetation productivity, as seen around the Andes.
Hence, it is imperative to inspect trends in GPP and negative carbon cycle extremes simultaneously to understand the nature of changes in extremes and their implications at regional scales.

We found that the duration of GPP TCEs increased with higher CO\textsubscript{2} concentration, i.e., a location experienced GPP TCEs with increasing duration over time.
The duration and impact of negative TCEs in GPP are enhanced when
human-induced LULCC was considered, resulting in increased loss in carbon uptake.
We illustrated this with the case study of the Chaco Province in
Argentina that had a net gain in carbon uptake (7.02~TgC) during
2000--24 in the simulation \emph{without LULCC} and while a net loss in
carbon uptake ($-$11.99~TgC) in the simulation \emph{with LULCC}.

The single most dominant climate driver was soil moisture that had highest correlations with GPP extremes ($p < 0.05$) and the correlations were mostly positive, indicating that anomalous decrease in soil moisture or drier climate conditions cause anomalous reductions in GPP.
Other major individual drivers were hot temperatures and fire.
Fire was a dominant climate driver, especially after 2100 in the
simulation \emph{with LULCC}, highlighting the increased vulnerability of ecosystems to fire events due to the impact of human activities on ecosystems.
We also found that the decline in precipitation triggers a negative
carbon cycle TCE event, but the reduction in soil moisture or water
limitation was the dominant driver for those negative carbon cycle TCE
events to persist.
The compound effects of climate drivers were also analyzed, and an
increasing number of regions under carbon cycle extremes were attributed to hot and dry conditions.
The largest fraction of negative carbon cycle extremes were driven by the compound effects of hot, dry, and fire events.
Warmer conditions under climate change increases the risk of occurrence of fire events and their impacts on the carbon cycle and extremes of the future.

This study presents a detailed analysis of detection and identification of carbon cycle extreme events, how these extremes evolve from 1850 through 2300, and how human-induced LULCC alters them using a fully coupled Earth system model forced with the RCP~8.5 and ECP~8.5 CO\textsubscript{2} concentration scenarios.
It also attributes the climate drivers of such extremes in carbon cycle
for the periods 1850--2100 and 2100--2300, and analyzes the changing
patterns and dominance of climate drivers, under \emph{with} and
\emph{without LULCC} scenarios.
Study provides new insights into the contribution of human activities in altering carbon cycle extremes and the vulnerability of terrestrial vegetation and associated ecosystem services that could present increasing risks to human lives, wildlife, and food security.

\newpage

\clearpage

\acknowledgments
We thank the Community Earth System Model (CESM)
Biogeochemistry Working Group (BGCWG) for performing the CESM1(BGC)
simulations analyzed in this study.
\noindent This research was supported by the Reducing Uncertainties in Biogeochemical Interactions through Synthesis and Computation (RUBISCO) Science Focus Area, which is sponsored by the Regional and Global Model Analysis (RGMA) activity of the Earth \& Environmental Systems Modeling (EESM) Program in the Earth and Environmental Systems Sciences Division (EESSD) of the Office of Biological and Environmental Research (BER) in the US Department of Energy Office of Science.
This research used resources of the Compute and Data Environment for Science (CADES) at the Oak Ridge National Laboratory, which is managed by UT-Battelle, LLC, for the US Department of Energy under Contract No.\ DE-AC05-00OR22725.

\section*{Open Research}

\noindent \textbf{Data Availability Statement}

\noindent The CESM1(BGC) model output \citep{Sharma_Data} used for detection and attribution of carbon cycle extremes in the study are available at \href{https://doi.org/10.5281/zenodo.5548153}{https://doi.org/10.5281/zenodo.5548153}.
\noindent Data analysis was performed in Python, and all analysis codes
are publicly available on GitHub \citep{Sharma_Codes} at
\href{https://github.com/sharma-bharat/Codes\_Carbon\_Extremes\_2300}{https://github.com/sharma-bharat/Codes\_Carbon\_Extremes\_2300}
and archieved at \href{https://doi.org/10.5281/zenodo.6147120}{https://doi.org/10.5281/zenodo.6147120}.

This paper was part of PhD Thesis which led to several papers \citep{warner,sharma_bg,sharma_icdm,shaema_jgrb,sharma2022analysis}.


%
%
\clearpage
\bibliography{bib_paper}
\clearpage

\section*{Supplementary Material}
\vspace{1cm}
\renewcommand{\thesection}{S\arabic{section}}
\renewcommand{\thetable}{S\arabic{table}}
\renewcommand{\thefigure}{S\arabic{figure}}
\setcounter{section}{0}
\setcounter{table}{0}
\setcounter{figure}{0}

\clearpage

\begin{figure}
 \centering
 \includegraphics[width=0.95\columnwidth]{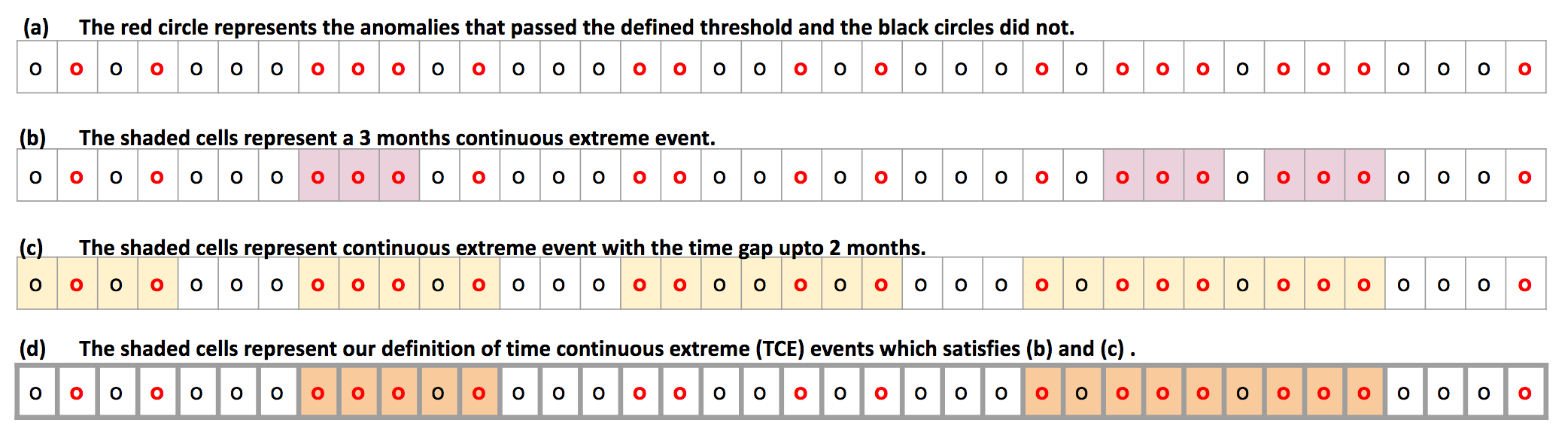}
 \caption{The schematic flow-diagram for finding a temporally contiguous extreme (TCE) event from a time series of variable anomalies at any grid cell. The black and red colored circles represent the anomalies that have not~passed and passed the thresholds, respectively. Hence, red circles are individual extreme months in a time series of a variable anomalies (a). We then look for 3~month continuous extremes (b) which is the first condition to qualify as a TCE event. We also look for the individual or temporally continuous extremes that are located in vicinity of each other up to 2 months (c). The temporally contiguous extreme events that fulfil both conditions (shown in (b) and (c)) are referred to as TCE events. }
\label{f:tce_schematic}
\end{figure}
 
\begin{figure}
 \centering
 \includegraphics[trim = {.1cm .1cm .1cm .8cm}, clip, width=0.85\columnwidth]{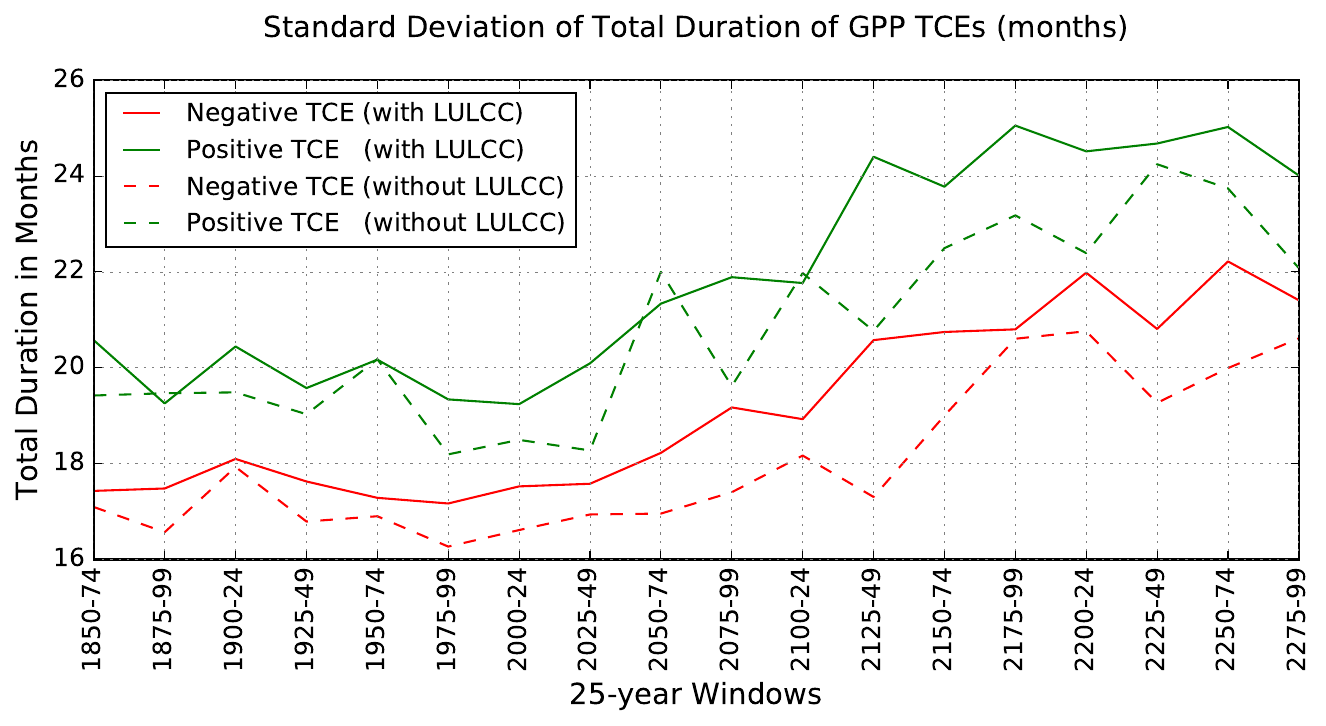}
 \caption{The standard deviation duration of temporally continuous extreme (TCE) events for every time window from 1850--2299. The figure shows the development of standard deviation duration of positive (shown in green) and negative (shown in red) TCEs for both the simulations, \emph{with} (solid lines) and \emph{without LULCC} (dashed lines).}
\label{f:ts_tce_len_std}
\end{figure}

\begin{figure}
 \centering
 \includegraphics[trim = {0.1cm 0.1cm .1cm .65cm},clip,width=0.95\columnwidth]{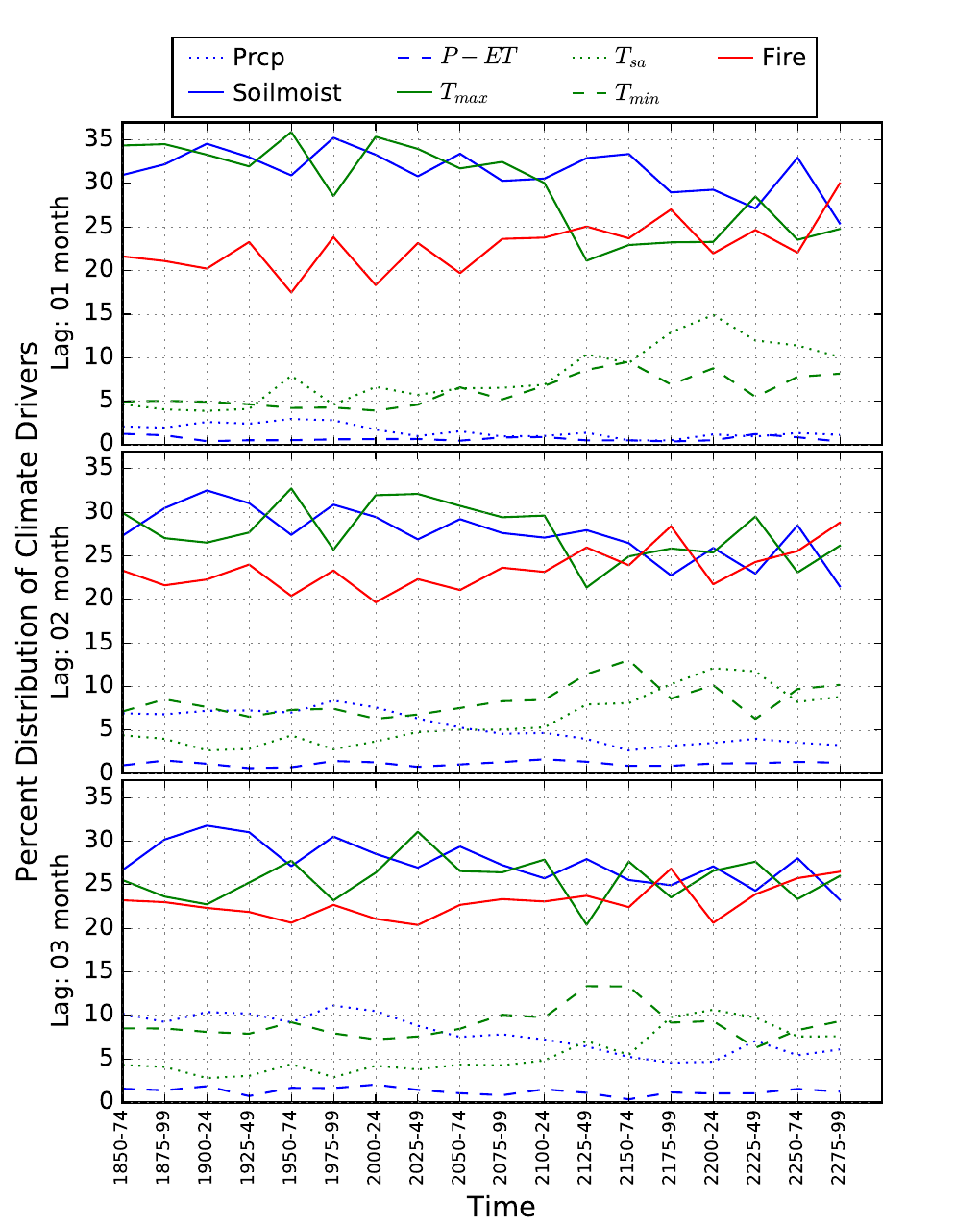}
 \caption{Percent distribution of global dominant climate drivers \emph{without LULCC} for every time window from 1850--2299. 
 For a particular lag month (1, 2, 3, etc.), a climate driver with highest correlation coefficient ($p$\,$<$\,$0.05$) with carbon cycle TCEs at any grid cell is called a dominant climate driver.}
\label{f:per_dom_wo_lulcc}
\end{figure}
 

\begin{figure}
 \subfloat[Precipitation]{\includegraphics[trim = {.4cm .1cm 2.8cm .6cm}, clip, width=0.5\columnwidth]{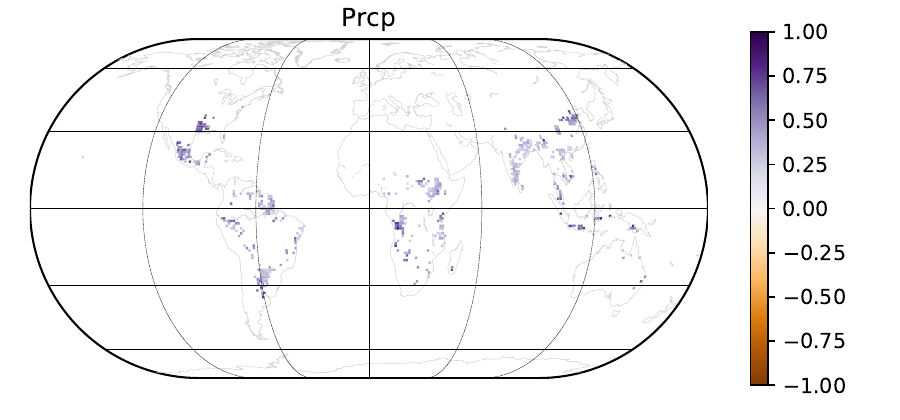} \label{f:rgb_neg_w_lulcc_w6l1_prcp}}
 \subfloat[Soil Moisture]{\includegraphics[trim = {.4cm .1cm 2.8cm .6cm}, clip, width=0.5\columnwidth]{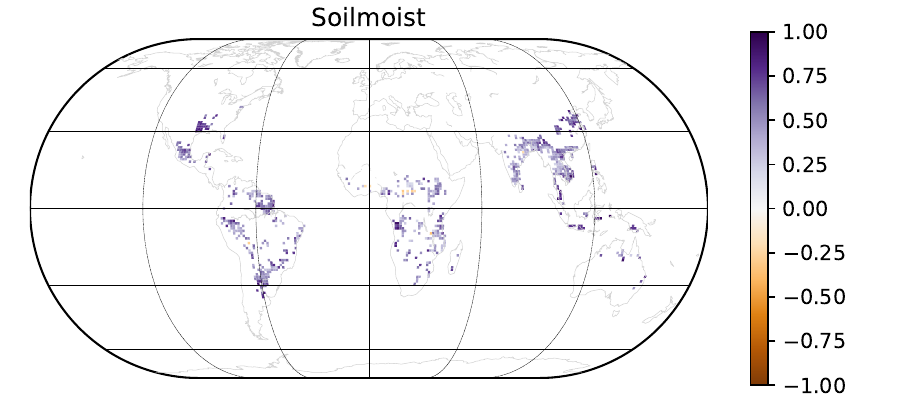} \label{f:rgb_neg_w_lulcc_w6l1_sm}} \\
 \begin{center}
 \subfloat[Correlation Coefficient]{\includegraphics[trim = {12.4cm .1cm .4cm .1cm}, clip, angle=270, width=5cm]{figures/rgb/neg_w_lulcc_based_pos_wo_lulcc/dri_win_06_lag_01_sm.pdf} \label{f:rgb_neg_w_lulcc_w6l1_cor}}    
 \end{center} 
 \subfloat[Monthly Max Temperature]{\includegraphics[trim = {.4cm .1cm 2.8cm .6cm}, clip, width=0.5\columnwidth]{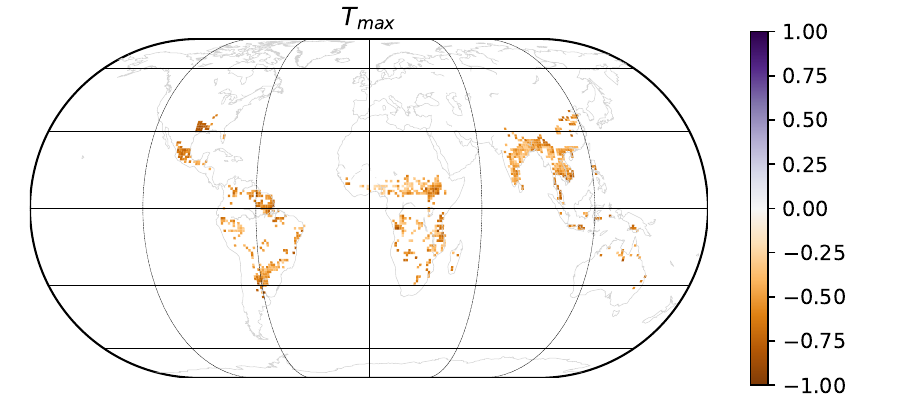} \label{f:rgb_neg_w_lulcc_w6l1_tmax}}
 \subfloat[Fire]{\includegraphics[trim = {.4cm .1cm 2.8cm .6cm}, clip, width=0.5\columnwidth]{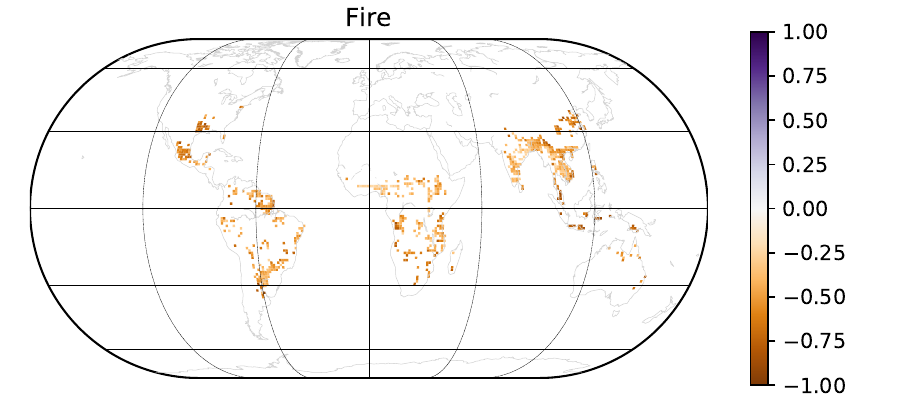} \label{f:rgb_neg_w_lulcc_w6l1_fire}}\\
 \caption{Spatial distribution of climate drivers driving temporally continuous extremes in GPP at a lag of 1~month for the time window 2000--24 for \emph{with LULCC}. Large losses in carbon uptake or increase in negative extremes, and reduction of precipitation and soil moisture are positively correlated. And, increase in temperatures and fire are negatively correlated with negative extremes in GPP. The compound effect of these climate drivers are shown in RGB maps (Figure~7).}
 \label{f:rgb_neg_w_lulcc_dri_win6_lag1}
\end{figure}

\begin{figure}
 \centering
 \includegraphics[width=0.95\columnwidth]{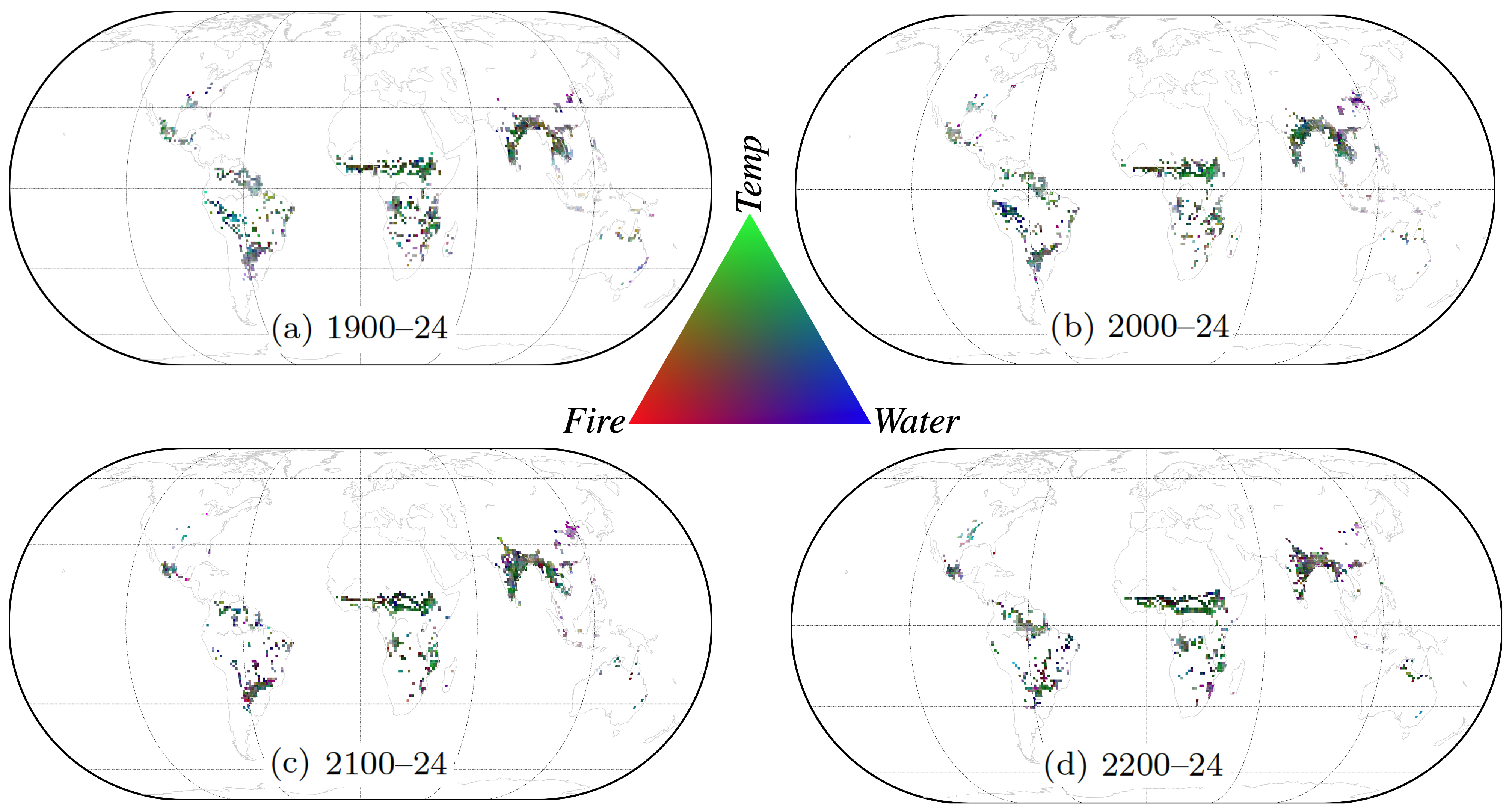}
 \caption{Spatial distribution of climate drivers attributing to negative TCEs for \emph{without LULCC} for four 25-year time windows, (a) 1900--24, (b) 2000--24, (c) 2100--24, and (d) 2200--24. The climate drivers are pooled in three colors, red, green, and blue. Red ($Fire$) is for loss of carbon due to fire, green ($Temp$) represents monthly maximum, mean, and minimum daily temperatures ($T_\mathrm{max}$, $T_\mathrm{sa}$, $T_\mathrm{min}$ respectively), Blue ($Water$) includes monthly means of soil moisture, precipitation and $P$$-$$E$ (precipitation minus evapotranspiration). The results shown here are at 1 month lag.}

\label{f:rgb_neg_wo_lulcc}
\end{figure}

\begin{figure}
 \centering
 \includegraphics[trim = {.08cm .1cm .1cm .95cm}, clip, width=0.85\columnwidth]{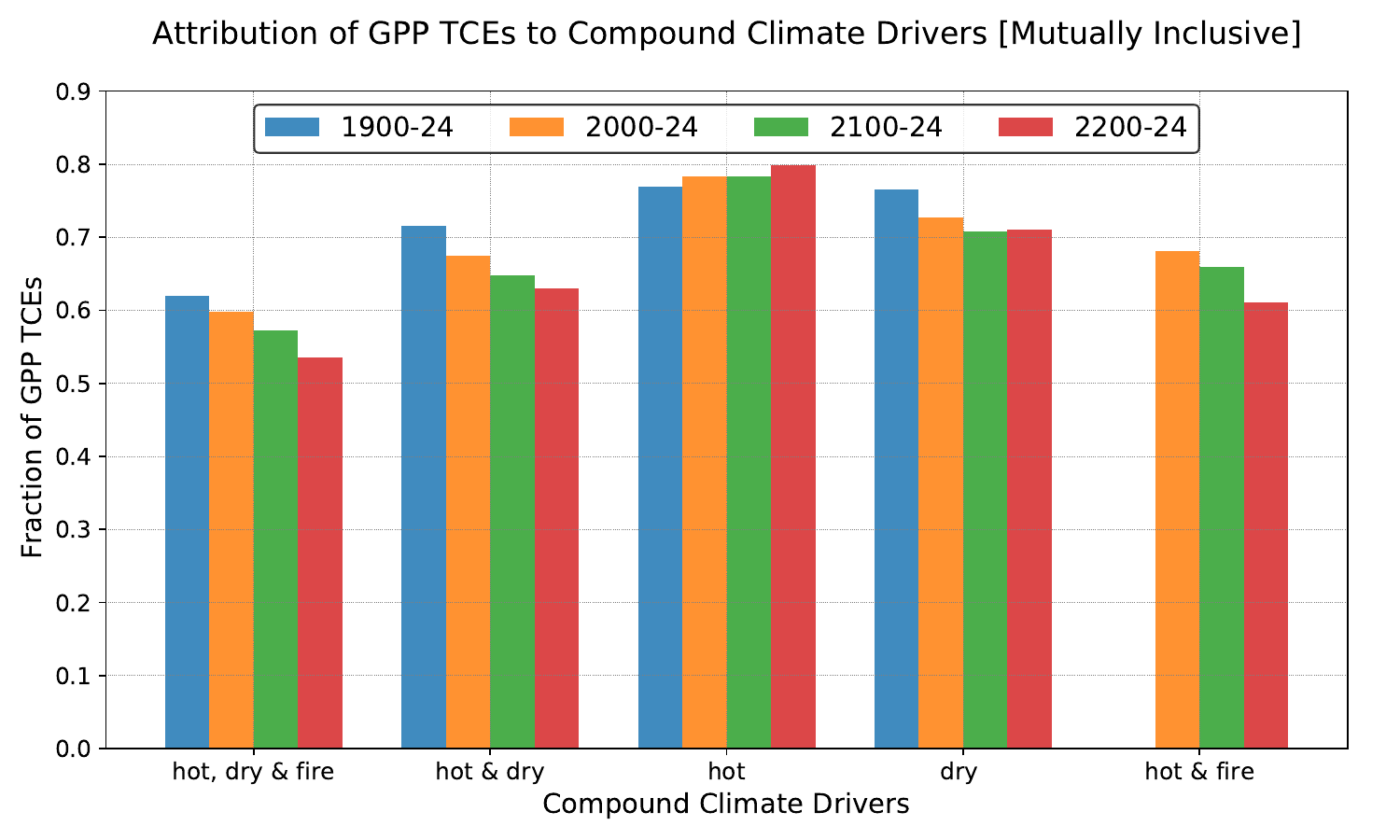}
 \caption{Attribution of temporally continuous extreme events in GPP to compound effect of climate drivers for \emph{with LULCC} at lag of 1~month for 25-year time windows, (a) 1900--24, (b) 2000--24, (c) 2100--24, and (d) 2200--24. The fractions are mutually inclusive, i.e., events driven by \textit{hot and dry} climate is also counted in either \textit{hot} or \textit{dry} climate driven events. Any location could be affected by one or compound climatic conditions. The chart here only represents the inclusiveness of the climatic conditions represented in Figure~8. The combined effect of hot and dry climate accompanied by fire leads to most negative TCE events in GPP.}
\label{f:rgb_neg_w_lulcc_frac_inclusive}
\end{figure}

\begin{figure}
 \subfloat[Change of Negative GPP Extremes (21\textsuperscript{st} century)]{\includegraphics[trim = {.5cm .1cm .7cm .1cm}, clip, width=0.5\columnwidth]{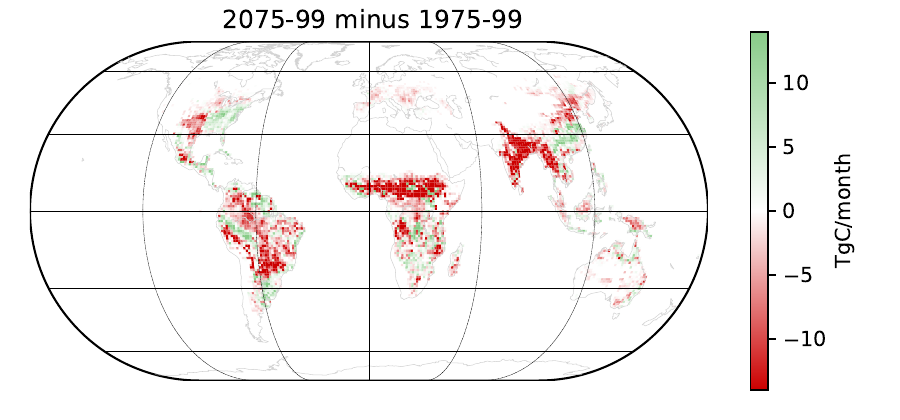}\label{f:negext_gpp_diff_21_wo_lulcc}}
 \subfloat[Change of total GPP (21\textsuperscript{st} century)]{\includegraphics[trim = {.5cm .1cm .7cm .1cm}, clip,width=0.5\columnwidth]{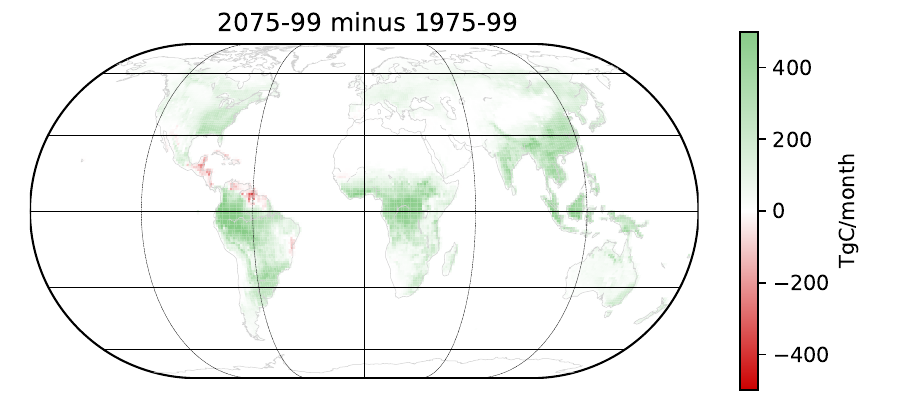}\label{f:gpp_diff_21_wo_lulcc}} \\
 \subfloat[Change of Negative GPP Extremes (23\textsuperscript{rd} century)]{\includegraphics[trim = {.5cm .1cm .7cm .1cm}, clip,width=0.5\columnwidth]{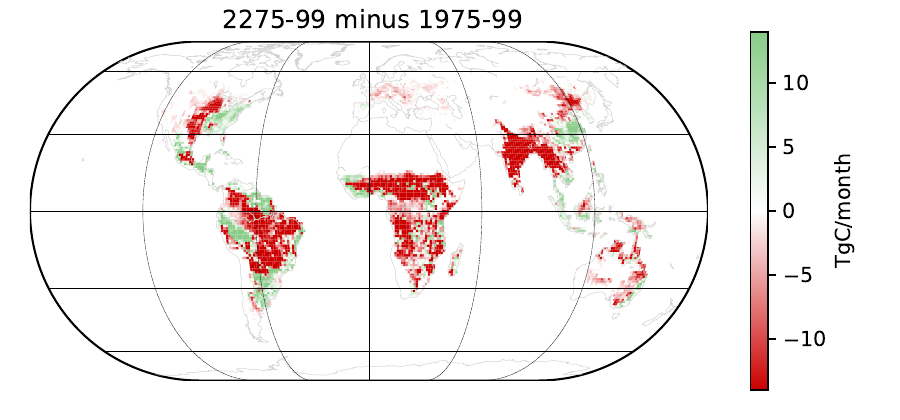}\label{f:negext_gpp_diff_23_wo_lulcc}}%
 \subfloat[Change of total GPP (23\textsuperscript{rd} century)]{\includegraphics[trim = {.5cm .1cm .7cm .1cm}, clip,width=0.5\columnwidth]{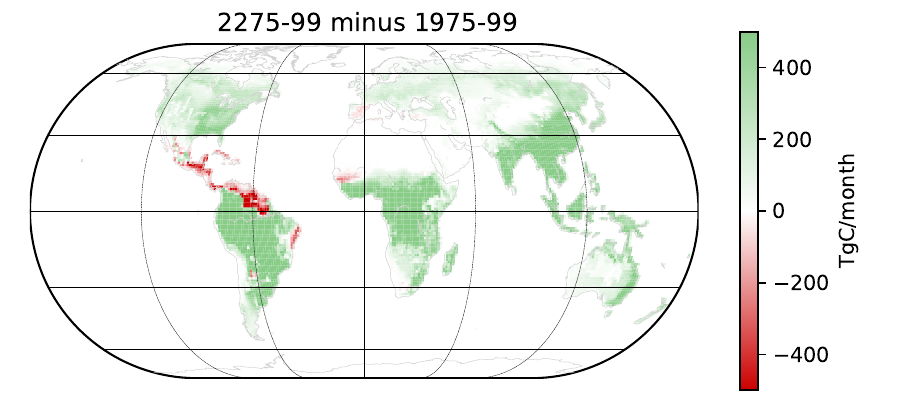}\label{f:gpp_diff_23_wo_lulcc}} \\
 \caption{The GPP negative extremes in \emph{without LULCC} (a)\&(c) and GPP (b)\&(d) are integrated over the whole globe and 25-year time periods. Red and green color in (a)\&(c) indicates the increasing and weakening intensity of negative extremes respectively. Red and green color in (b)\&(d) indicates the loss and increase of vegetation respectively.}
 \label{f:diff_gpp_and_negext_wo_lulcc}
\end{figure}


\begin{figure}
 \subfloat[The IAV of GPP.]{\includegraphics[trim = {0.1cm 2.2cm 0.1cm .85cm}, clip, width=0.5\columnwidth]{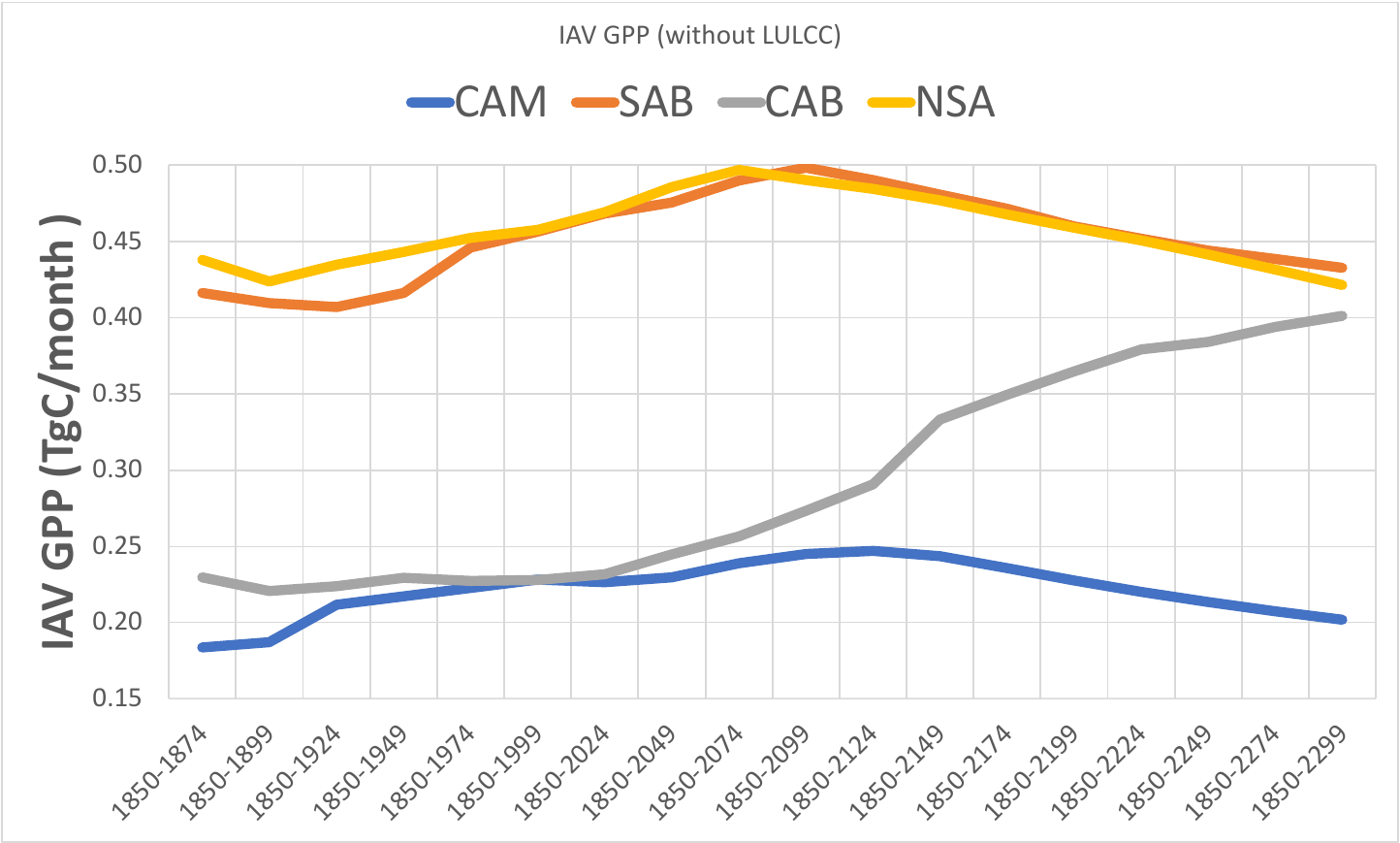}\label{f:gpp_regions_wo_iav}} \\
 \vspace{-3.5cm}
 \subfloat[Changes in GPP.]{\includegraphics[trim = {0.1cm 0.1cm 0.1cm 2.1cm}, clip,width=0.5\columnwidth]{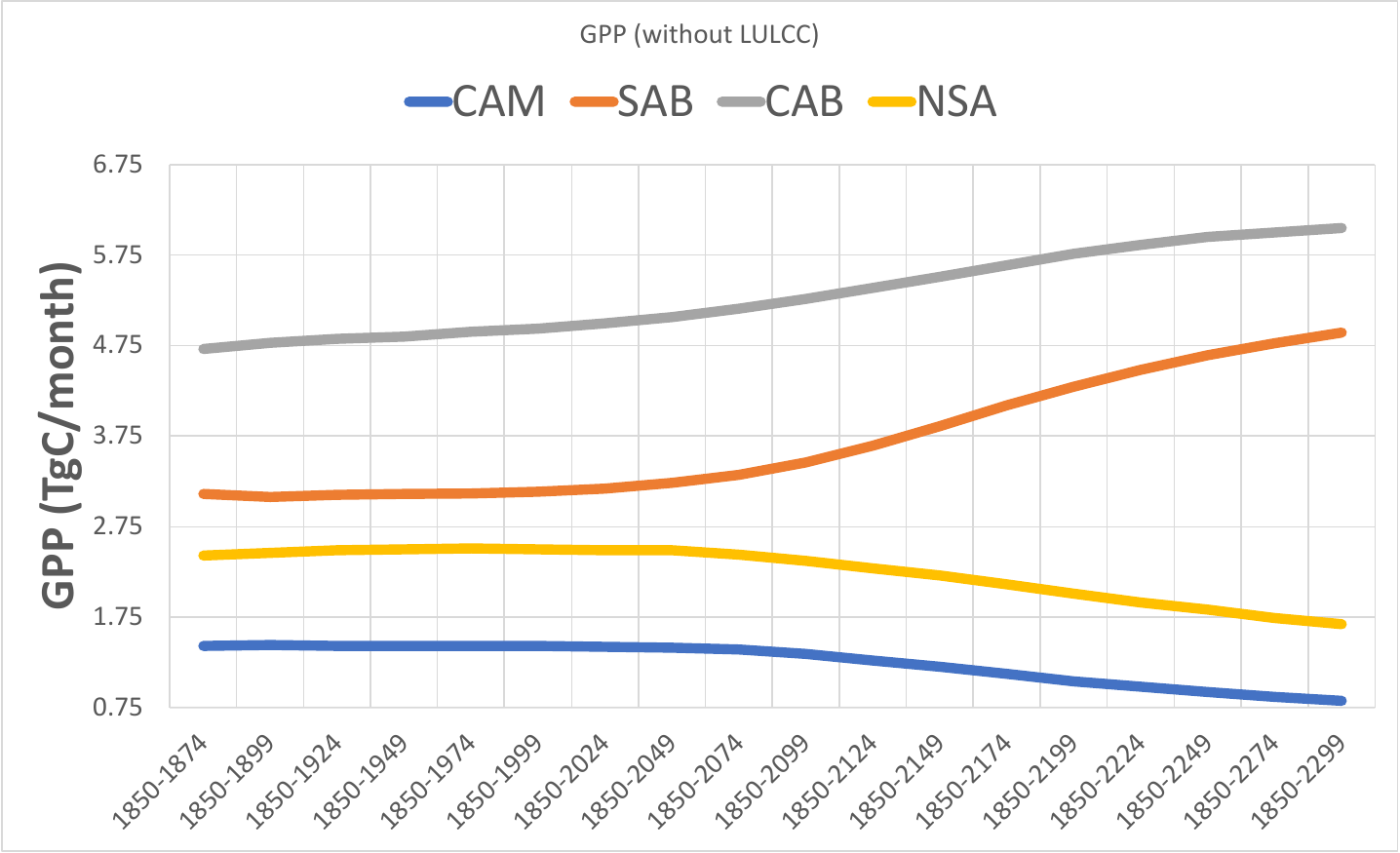}\label{f:gpp_regions_wo_trend}} 
 \subfloat[The regions of interest in the Americas.]{\includegraphics[trim = {0.1cm 0.1cm 0.1cm .85cm}, clip, width=0.55\columnwidth]{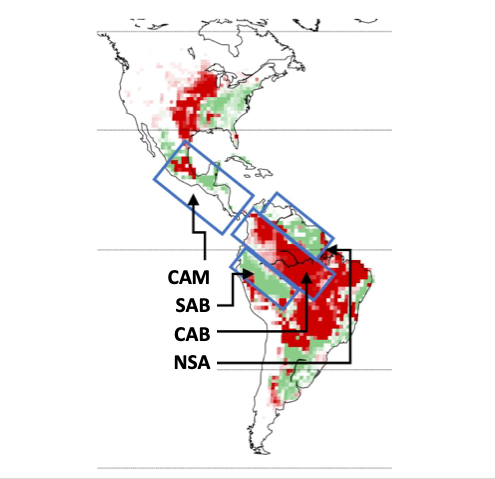}\label{f:discuss_regions_diff_neg_ext}} \\
 
 \subfloat[Changes in precipitation.]{\includegraphics[trim = {0.1cm 0.1cm 0.1cm .85cm}, clip,width=0.5\columnwidth]{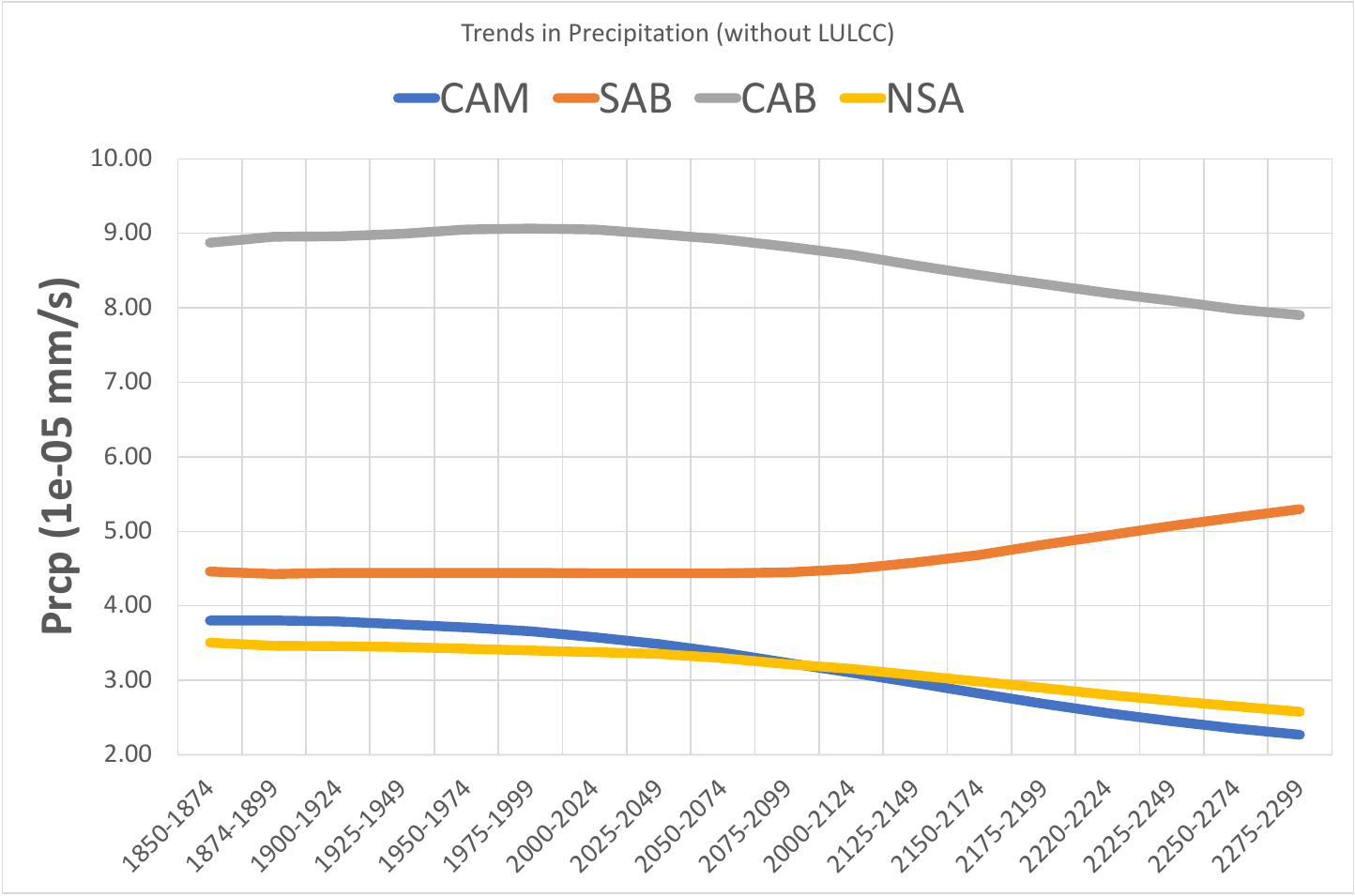}\label{f:prcp_regions_wo_trend}}
 \subfloat[Changes in soil moisture.]{\includegraphics[trim = {0.1cm 0.1cm 0.1cm .85cm}, clip,width=0.5\columnwidth]{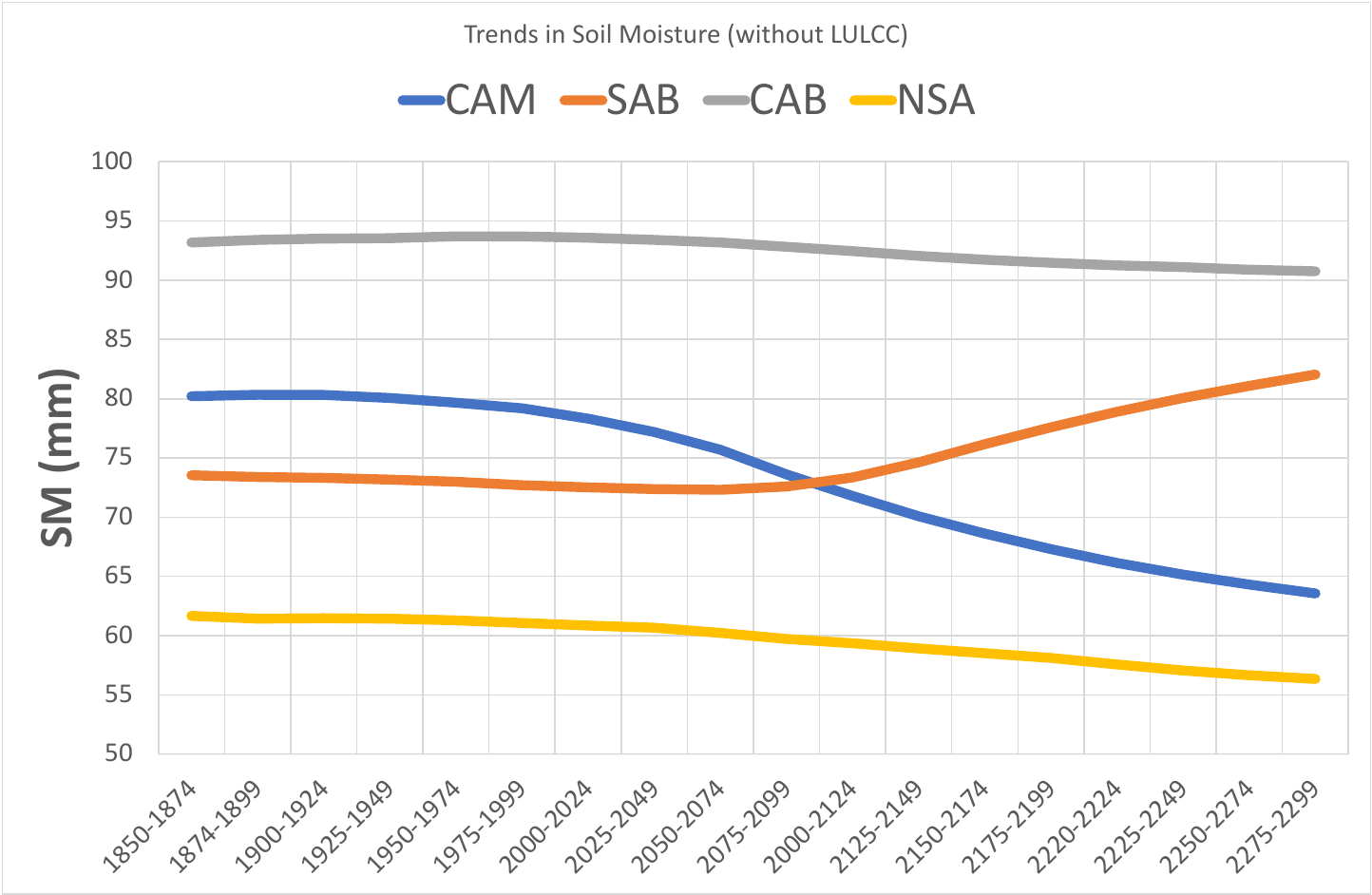}\label{f:sm_regions_wo_trend}}\\
 
 \caption{The IAV of GPP and changes in GPP and climate drivers are shown for \emph{without LULCC}. 
 The IAV and changes are calculated from 1850 as the base year to 25~year increments, as shown in $x$-axis.}
 \label{f:climate_drivers_disc}
\end{figure}

\begin{figure}                                                                  
 \centering                                                                     
 \subfloat[Change of Negative GPP Extremes (23\textsuperscript{rd} century)]{\includegraphics[trim = {.5cm .1cm .7cm .1cm}, clip, width=0.5\columnwidth]{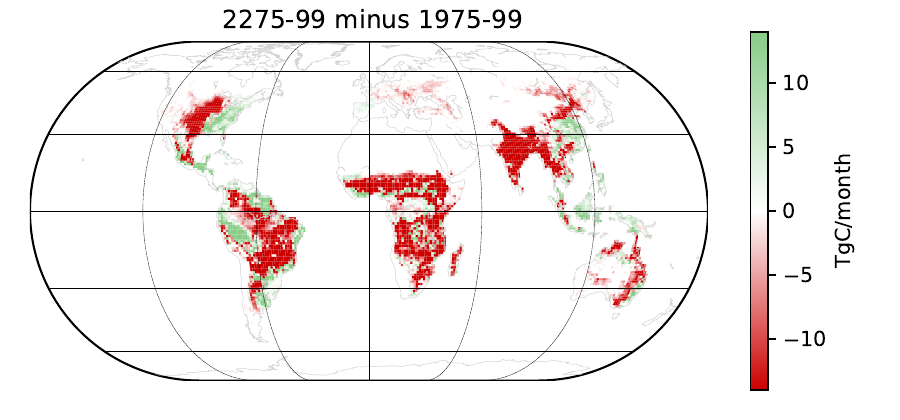}\label{f:negext_gpp_diff_23_w_lulcc}}%
 \subfloat[Change of total GPP (23\textsuperscript{rd} century)]{\includegraphics[trim = {.5cm .1cm .7cm .1cm}, clip, width=0.5\columnwidth]{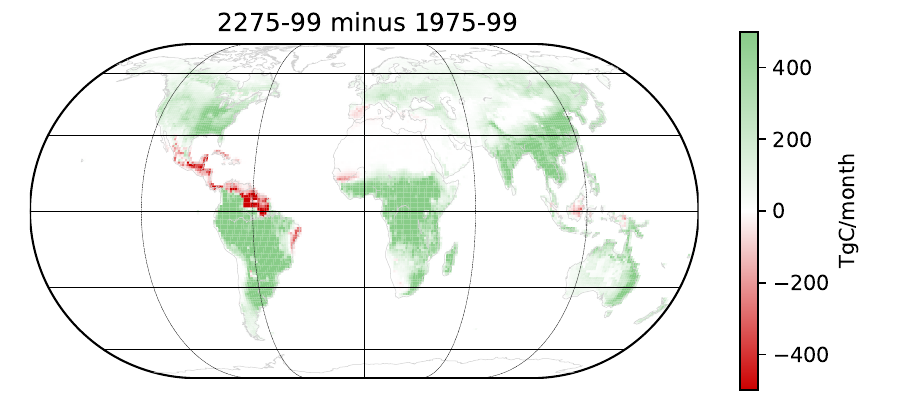}\label{f:gpp_diff_23_w_lulcc}} \\
                                                                                 
 \caption{The GPP negative extremes in \emph{with LULCC} (a) and GPP (b) are integrated over the whole globe and 25-year time period. Red  and green color in (a) indicates the increasing and weakening intensity of negative extremes respectively. Red and green color in (b) indicates the loss     and increase of vegetation respectively. The patterns are similar to the \emph{without LULCC} except for Indonesia which shows the decline of productivity and weakening of negative extreme events.}
  \label{f:diff_gpp_and_negext_w_lulcc}                                         
 \end{figure}   
\begin{figure}
 \centering
 \subfloat[Changes in Soil Moisture (21\textsuperscript{st} century)]{\includegraphics[trim = {.8cm .1cm .8cm 1.2cm}, clip, width=0.5\columnwidth]{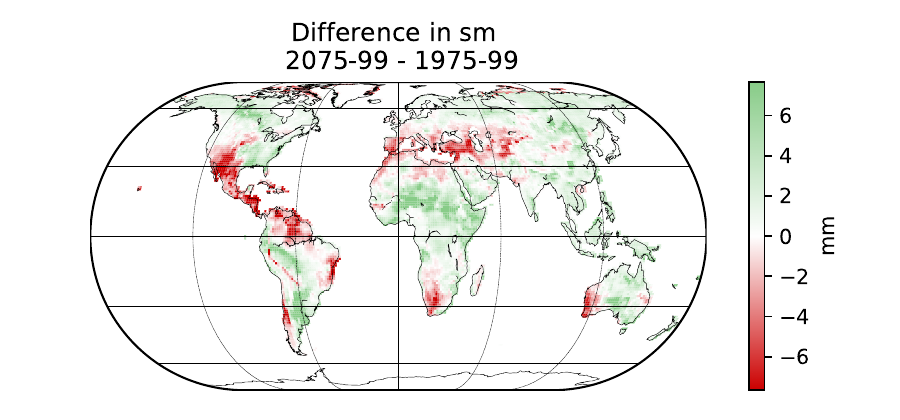}\label{f:sm_diff_21_wo_lulcc}}%
 \subfloat[Changes in Soil Moisture (23\textsuperscript{rd} century)]{\includegraphics[trim = {.8cm .1cm .8cm 1.2cm}, clip, width=0.5\columnwidth]{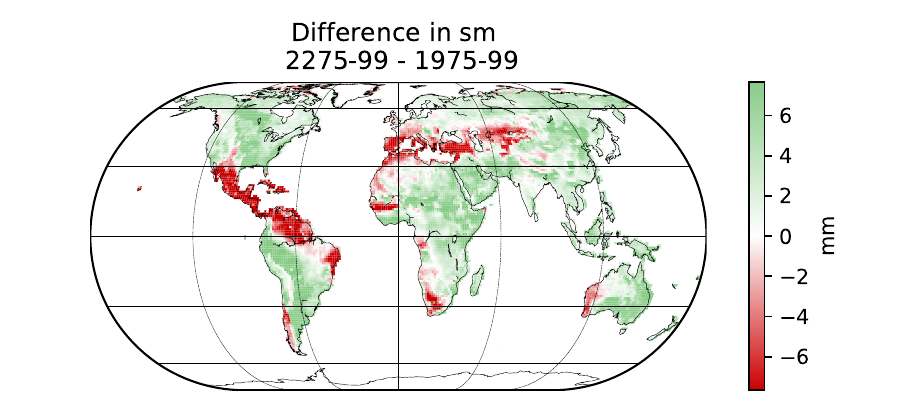}\label{f:sm_diff_23_wo_lulcc}} \\
 
 \caption{The area weighted average of Soil Moisture for \emph{without LULCC}. (a) Changes in soil moisture for years 2075--2099 minus 1975--1999 and (b) changes in soil moisture for years 2275--2099 minus 1975--1999. Red and green color in indicates the reduction and increase of soil moisture respectively.}
  \label{f:diff_sm_wo_lulcc}
\end{figure}

\begin{figure}
 \subfloat[\emph{ without LULCC}]{\includegraphics[trim={0.1cm 0.1cm 0.1cm .85cm},clip,width=0.95\columnwidth]{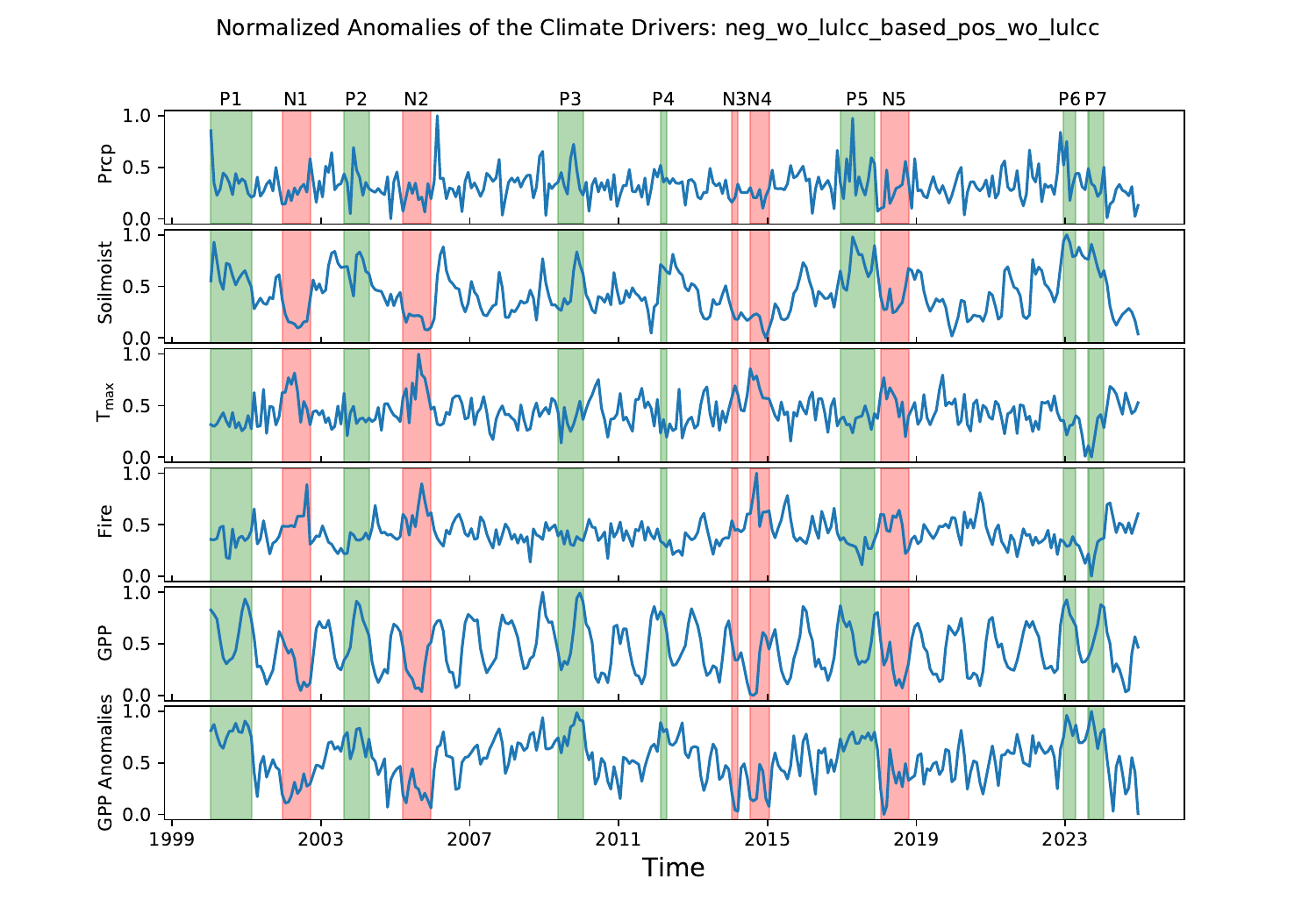}}\label{f:ts_tce_len_mean}
 
 \subfloat[\emph{with LULCC}]{\includegraphics[trim={0.1cm 0.1cm 0.1cm .85cm},clip,width=0.95\columnwidth]{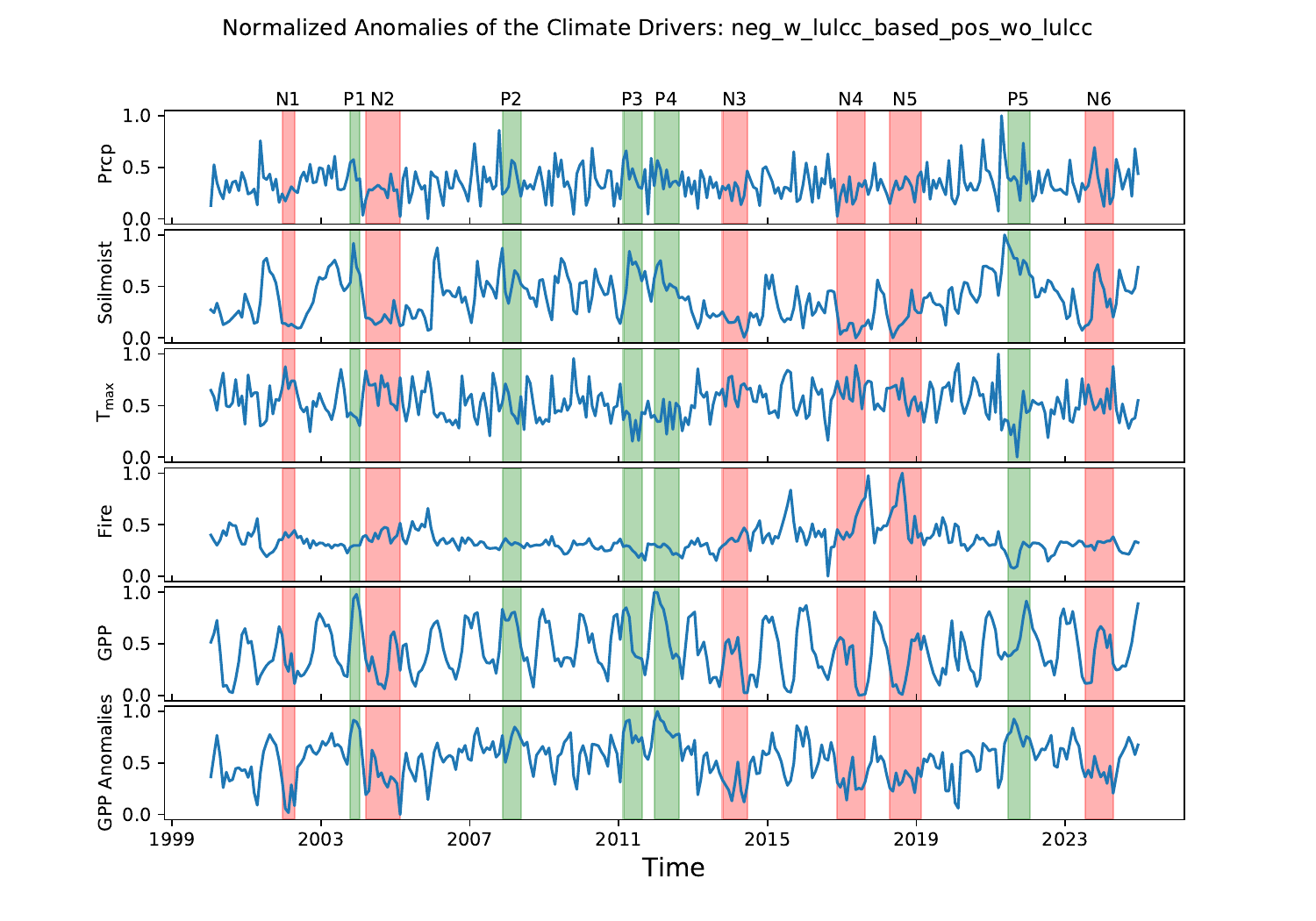}}\label{f:tce_kde_neg_len}
 
 \caption{Time series of normalized anomalies of climate drivers and GPP at Chaco Province, Argentina for (a) \emph{ without LULCC} and (b) \emph{ with LULCC} during 2000--24.
 The shaded areas in green color span over the positive TCEs ((a): P1 to P7 and (b): P1 to P5).
 Similarly, the areas in red color represents the negative GPP TCEs ((a): N1 to N5 and (b): N1 to N6).}
 \label{f:tce_wo_lulcc_dri_ano_ag_combined}
\end{figure}

\begin{table}[]
\centering
\caption{ILAMB score of the variables of CESM1(BGC)}{Source: \url{https://www.ilamb.org/CMIP5/historical/}}
\label{t:table_ilamb}
\begin{tabular}{|l|l|l|}
\hline
\textbf{Variable/Benchmark Data}                        & \textbf{IAV Score} & \textbf{Overall Score} \\ \hline
Surface Air Temperature/CRU              & 0.821              & 0.782                  \\ \hline
Diurnal Max Temperature/CRU             & 0.793              & 0.752                  \\ \hline
Precipitation/ GPCC                       & 0.793              & 0.651                  \\ \hline
Terrestrial Water Storage Anomaly/GRACE & 0.905              & 0.588                  \\ \hline
Burned Area/GFED4S                       & 0.447              & 0.409                  \\ \hline
Evapotranspiration/MODIS                 & 0.617              & 0.588                  \\ \hline
\end{tabular}
\end{table}

\begin{table}
\centering
\caption{Details of GPP TCEs and PFT distribution at Chaco Province, Argentina. The results are shown for \emph{without LULCC} and time window 2000--24. PFT refers to plant functional type, BDT is broadleaf deciduous tree, and BET is broadleaf evergreen tree.}
\label{t:tce_wo_lulcc_ag}
\resizebox{\textwidth}{!}{%
\begin{tabular}{|c|c|c|c|c|}
\hline
\textbf{TCE - neg}     & \textbf{TCE - pos}      & \textbf{TCE length - neg} & \textbf{TCE length - pos} & \textbf{TCE length - total} \\ \hline
5 (events)             & 7 (events)              & 40 (months)               & 53 (months)               & 93 (months)                 \\ \hline
\hline
\textbf{PFT (I)}       & \textbf{PFT(II)}        & \textbf{PFT(III)}         & \textbf{Latitude}         & \textbf{Longitude}          \\ \hline
BDT Temperate (43.2\%) & BET Temperate (17.91\%) & C$_3$ grass (17.48\%)        & 25.916$^{\circ}$ S                 & 300$^{\circ}$ E                      \\ \hline
\end{tabular}%
}
\end{table}

\begin{table}[]
\centering
\caption{Linear regression results for attribution analysis using the cumulative lagged effects for the region of Chaco Province, Argentina. The results are shown for \emph{without LULCC} and time window 2000--24.}
\label{t:reg_wo_lulcc_win6_cum_lag}
\resizebox{\textwidth}{!}{%
\begin{tabular}{|c|c|c|c|c|c|c|c|c|c|c|c|c|c|c|}
\hline
\textbf{} & \multicolumn{2}{c|}{Fire} & \multicolumn{2}{c|}{$P-E$} & \multicolumn{2}{c|}{Precipitation} & \multicolumn{2}{c|}{Soil Moisture} & \multicolumn{2}{c|}{$T_\mathrm{max}$} & \multicolumn{2}{c|}{$T_\mathrm{min}$} & \multicolumn{2}{c|}{$T_\mathrm{sa}$} \\ \hline
\textbf{Lags} & CC & PV & CC & PV & CC & PV & CC & PV & CC & PV & CC & PV & CC & PV \\ \hline
\textbf{1} & -0.65 & 1.76E-12 & 0.365 & 3.27E-04 & 0.557 & 6.56E-09 & 0.734 & 5.85E-17 & -0.677 & 8.89E-14 & -0.328 & 1.31E-03 & -0.576 & 1.54E-09 \\ \hline
\textbf{2} & -0.653 & 1.25E-12 & 0.479 & 1.16E-06 & 0.653 & 1.25E-12 & 0.678 & 8.61E-14 & -0.693 & 1.42E-14 & -0.35 & 5.86E-04 & -0.592 & 4.29E-10 \\ \hline
\textbf{3} & -0.647 & 2.44E-12 & 0.502 & 2.99E-07 & 0.669 & 2.24E-13 & 0.639 & 5.61E-12 & -0.676 & 1.02E-13 & -0.364 & 3.36E-04 & -0.585 & 7.16E-10 \\ \hline
\textbf{4} & -0.624 & 2.45E-11 & 0.4918 & 5.53E-07 & 0.673 & 1.50E-13 & 0.621 & 3.26E-11 & -0.674 & 1.37E-13 & -0.36 & 3.96E-04 & -0.584 & 7.97E-10 \\ \hline
\end{tabular}%
}
\end{table}

\begin{table}[]
\centering
\caption{Linear regression results for attribution to climate driver triggers (i.e. onset 25\% of TCE length) and cumulative lagged effects for the region of Chaco Province, Argentina. 
The results are shown for \emph{without LULCC} and time window 2000--24.}
\label{t:reg_wo_lulcc_win6_cum_lag_trig}
\resizebox{\textwidth}{!}{%
\begin{tabular}{|c|c|c|c|c|c|c|c|c|c|c|c|c|c|c|}
\hline
& \multicolumn{2}{c|}{Fire} & \multicolumn{2}{c|}{$P-E$} & \multicolumn{2}{c|}{Precipitation} & \multicolumn{2}{c|}{Soil Moisture} & \multicolumn{2}{c|}{$T_\mathrm{max}$} & \multicolumn{2}{c|}{$T_\mathrm{min}$} & \multicolumn{2}{c|}{$T_\mathrm{sa}$} \\ \hline
\textbf{Lags} & CC & PV & CC & PV & CC & PV & CC & PV & CC & PV & CC & PV & CC & PV \\ \hline
\textbf{1} & -0.679 & 5.08E-05 & 0.557 & 1.68E-03 & 0.718 & 1.16E-05 & 0.609 & 4.49E-04 & -0.622 & 3.12E-04 & -0.331 & 7.93E-02 & -0.534 & 2.84E-03 \\ \hline
\textbf{2} & -0.602 & 5.50E-04 & 0.595 & 6.57E-04 & 0.697 & 2.71E-05 & 0.469 & 1.03E-02 & -0.597 & 6.32E-04 & -0.259 & 1.75E-01 & -0.494 & 6.49E-03 \\ \hline
\textbf{3} & -0.518 & 3.98E-03 & 0.569 & 1.29E-03 & 0.662 & 9.15E-05 & 0.361 & 5.46E-02 & -0.527 & 3.35E-03 & -0.207 & 2.81E-01 & -0.424 & 2.19E-02 \\ \hline
\textbf{4} & -0.377 & 4.39E-02 & 0.439 & 1.71E-02 & 0.605 & 5.11E-04 & 0.353 & 6.05E-02 & -0.479 & 8.51E-03 & -0.125 & 5.19E-01 & -0.36 & 5.53E-02 \\ \hline
\end{tabular}%
}
\end{table}

\begin{table}[]
\centering
\caption{Linear regression results for attribution to climate driver triggers (i.e. onset 25\% of TCE length) and cumulative lagged effects for the region of Chaco Province, Argentina. 
The results are shown for the simulation \emph{with LULCC} and time window 2000--24.}
\label{t:reg_w_lulcc_win6_cum_lag_trig}
\resizebox{\textwidth}{!}{%
\begin{tabular}{|c|c|c|c|c|c|c|c|c|c|c|c|c|c|c|}
\hline
 & \multicolumn{2}{c|}{Fire} & \multicolumn{2}{c|}{$P-E$} & \multicolumn{2}{c|}{Precipitation} & \multicolumn{2}{c|}{Soil Moisture} & \multicolumn{2}{c|}{$T_\mathrm{max}$} & \multicolumn{2}{c|}{$T_\mathrm{min}$} & \multicolumn{2}{c|}{$T_\mathrm{sa}$} \\ \hline
\textbf{Lags} & CC & PV & CC & PV & CC & PV & CC & PV & CC & PV & CC & PV & CC & PV \\ \hline
\textbf{1} & -0.457 & 1.69E-02 & 0.574 & 1.74E-03 & 0.668 & 1.37E-04 & 0.549 & 3.00E-03 & -0.73 & 1.53E-05 & 0.062 & 7.57E-01 & -0.578 & 1.59E-03 \\ \hline
\textbf{2} & -0.353 & 7.12E-02 & 0.598 & 9.90E-04 & 0.614 & 6.47E-03 & 0.428 & 2.59E-02 & -0.581 & 1.49E-03 & 0.056 & 7.81E-01 & -0.418 & 2.90E-02 \\ \hline
\textbf{3} & -0.236 & 2.37E-01 & 0.502 & 7.60E-03 & 0.512 & 6.30E-03 & 0.376 & 5.33E-02 & -0.357 & 6.75E-02 & 0.108 & 5.92E-01 & -0.186 & 3.54E-01 \\ \hline
\textbf{4} & -0.2 & 3.16E-01 & 0.416 & 3.09E-02 & 0.454 & 1.73E-02 & 0.367 & 5.90E-02 & -0.293 & 1.38E-01 & 0.07 & 7.28E-01 & -0.12 & 5.50E-01 \\ \hline
\end{tabular}%
}
\end{table}





\end{document}